\documentclass[prb,twocolumn,showpacs,amsmath,amssymb,floatfix]{revtex4}
\usepackage{graphicx}
\usepackage{graphics}
\usepackage{dcolumn}
\usepackage{bm}
\usepackage{amsmath}
\usepackage{latexsym}
\usepackage{amsfonts}
\usepackage{amssymb}
\usepackage[rightcaption]{sidecap}
\usepackage{color}
\usepackage{float}
\usepackage{topcapt}
\usepackage{multirow}
\usepackage{yfonts}
\makeatletter
\newcommand{\mprb}{Phys. Rev. B}

\newcommand{\Rmnum}[1]{\expandafter\@slowromancap\romannumeral  #1@}
\def\bH{\textbf{H}}
\def\bU{\textbf{W}}
\def\Q{\Psi}
\def\bQ{\textbf{B}}
\def\ket{\rangle}
\def\q{\psi}
\def\bra{\langle}
\def\a{\alpha}
\def\b{\beta}
\def\rd{{\rm d}}
\def\d{\delta}
\def\m{\zeta}
\def\s{\gamma}
\newcommand{\eeq}{\end{eqnarray}}
\newcommand{\ee}{\end{equation}}
\newcommand{\beq}{\begin{eqnarray}}
\newcommand{\be}{\begin{equation}}

\newcommand{\fff}{\mbox{\scriptsize \boldmath $B$}}
\makeatother
\begin{document}
\title{Interacting fermions in 1D disordered lattices: \\Exploring localization and transport properties with lattice density-functional theories}
\date{\today}
\author{V. Vettchinkina}
\author{A. Kartsev}
\author{D. Karlsson}
\author{C. Verdozzi}
\affiliation{Mathematical Physics and European Theoretical Spectroscopy Facility (ETSF), Lund University, 22100  Lund, Sweden}%
\begin{abstract}
We investigate the static and dynamical behavior of 1D interacting fermions in disordered Hubbard chains,
contacted to semi-infinite leads. The chains are described via the repulsive Anderson-Hubbard 
Hamiltonian, using static and time-dependent lattice density-functional theory. The dynamical behavior of 
our quantum transport system is performed via an integration scheme available in the literature,
which we modify via the recursive Lanczos method, to increase its efficiency.
To quantify the degree of localization due to disorder and interactions, we adapt the definition of the inverse 
participation ratio
to obtain an indicator which is both suitable for quantum transport geometries and which can be obtained within density-functional theory.
Lattice density functional theories are reviewed and,
for contacted chains, we analyze the merits and limits of the coherent-potential approximation in 
describing the spectral properties, with interactions included via lattice density functional theory. 
Our approach appears to able
to capture complex features due to the competition between disorder and interactions.
Specifically, we find a dynamical enhancement of delocalization in presence of a finite bias,
and an increase of the steady-state current
induced by inter-particle interactions. 
This behavior is corroborated by results for the time-dependent densities and for the 
inverse participation ratio. Using short isolated chains with interaction and disorder, 
a brief comparative analysis between time-dependent density-functional theory and exact results
is then given, followed by general conclusive remarks.   
\end{abstract}

\maketitle
\section{Introduction}
In many physical phenomena, practical limitations hinder a complete knowledge of all the degrees of freedom involved. 
Nanoscience has adopted such apparent shortcoming as its central paradigm, by exploiting the notion of a small system 
coupled to a macroscopic environment. A case in point is represented by nanoscale transport phenomena, where two (or more)
macroscopic leads are connected to small central devices (quantum constriction) \cite{Datta,Ratner}. 

Such devices, whose size ranges from that of few atoms (as in short nanowires or 
small molecules) to that of several repeated large molecular units, attract scientific interest because they are seen as possible 
candidates for novel electronic, spintronic, or quantum computation devices, to mention
a few \cite{RecentQTreview}. This, in turn, requires a thorough understanding and control
of the decoherence processes which can affect carrier propagation and manipulation
in the device region. 

In this work we consider two of such processes, 
namely disorder and inter-particle interactions (thus leaving out other
important decoherence mechanisms, e.g., lattice vibrations). 
How interactions and disorder affect 
the conduction properties of materials has been intensively 
investigated over several decades
\cite{Anderson58,Mott,Lloyd,Thouless,CPA2,gang4,AHM, MacKinnoreview,Pendry,Gavish,Scalettar,eversmirlin,reviewdisorderinteraction},
and significant progress has been made. However, some
issues remain at a considerable extent open, e.g. the real-time dynamics
of samples with disorder and interactions.

Starting with the seminal paper by Anderson \cite{Anderson58}, lattice models have had an eminent place in the study of disordered systems with and without interactions. While a large fraction of the literature
on disordered interacting lattice models concentrates on the equilibrium regime (for both finite and extended systems), more recently the time-dependent properties have also been examined, primarily for finite samples \cite{fermionnegativepaper, Inguscio}. In between the finite/infinite-system categories, a third one is represented by small disordered samples connected to semi-infinite homogenous reservoirs \cite{Beenakker}, of relevance 
to quantum transport phenomena. 

This paper looks into some aspects of the transport properties of 1D interacting fermions in disordered lattice systems, using static \cite{staticDFT} and time-dependent density functional theory \cite{rg84} (DFT and TDDFT, respectively).
Static and time-dependent DFT are in principle exact reformulations of the (time-dependent)
many-body problem \cite{RvLproofs}, where the key variable is the one-particle density $n$, and a central ingredient is the 
exchange-correlation (XC) potential $v_{xc}$ (recent comprehensive reviews of the subject, are Refs. 
\onlinecite{TDDFTbook,TDDFTbook1,Ullrichbook}).
The XC potential embodies 
the complexities of the many-body problem. In this contracted description, $v_{xc}$ is a highly non-trivial functional of the density
(in TDDFT, where time enters explicitly into the formulation, such functional dependence
includes the entire history of the density $n$, i.e. memory effects). In general, the exact  $v_{xc}$
is not known, and approximations are introduced. A simple but not always adequate prescription is the so-called (adiabatic) local density approximation, where the XC potential depends only on the local (time-dependent) density. This amounts to neglecting non-locality in space (and memory effects in the TD case) in $v_{xc}$. As a result of this oversimplification,
in some practical applications an accurate description of dynamical inter-particle correlations may be lacking.

The application of static density-functional theories
to lattice models started almost thirty years ago\cite{GSDFT1,GSDFT1bis,GSDFT2,GSN},
and in the last decade
this approach has been further developed \cite{LimaPRL03,LimaEPL02,CapellePolini,Polinidisord}. The
use of lattice TDDFT to describe the non-equilibrium dynamics of Hubbard-like models 
is a rather new topic \cite{Baer08,Verdozzi08,Ullrich08,Polini08}, and 
some of its formal aspects are still under scrutiny (firm conceptual ground has been 
established for related lattice approaches which use the lattice bond-current 
as the basic variable \cite{GSEPMC10, KurthStefanucci2011,Tokatly11}). 
However, there is a growing body of evidence (see Sect. \ref{genaspDFT})
suggesting that the lack of a rigorous formulation (an issue which is likely to
be resolved in future) could be of no significant practical consequence.

After the above, somewhat lengthy, considerations, we can now define 
the motivations behind this work and the chosen methodology.  
Our focus is on finite chains contacted to semi-infinite leads, with
short-range interactions and disorder present only in the chains (the "device").
Even within these narrow boundaries, the issues which can potentially be addressed 
are many, but we will only touch upon a few of them, and going in no great detail.
In this respect, our work is somewhat exploratory in character, since we also
describe some methodological developments, that we found 
necessary when using (TD)DFT for disordered
lattice systems in the quantum transport regime. Concerning
the chosen methodology, we note that, compared to other lattice approaches,
lattice (TD)DFT is well suited for 
any dimensionality and is relatively inexpensive from the computational point of view
(since it deals with single-particle orbitals). These are attractive features when,
for example, one needs to perform configuration averages of time-evolved quantities
in the long-time limit (such as when approaching the steady state regime in
a quantum transport setup with disordered samples).

\indent The outline of the paper is as follows: 
In Sect.\ref{sec_hamilt}, we describe the lattice model system that we 
employ in our work. In Sect. \ref{genaspDFT}, we present the lattice (TD)DFT formalism, and describe how to obtain the XC potential from the exact solution of the 1D Hubbard model. This is accompanied by
a review of the inherent literature, to illustrate the
developments and applications occurred so far within this approach. 
In Sect. \ref{SectDis}, which deals with disorder, we first discuss the inverse participation ratio, then we introduce a
formulation for contacted chains based on the Coherent-Potential Approximation and DFT.
In Sect.\ref{TDQTinTDDFT} we start by briefly reviewing lattice TDDFT approaches to quantum transport. Then, we present in some detail a method recently proposed in the literature \cite{practical}, followed by a description of our modifications to it,  to increase its efficiency. 
Some technical details relative to Sects. \ref{SectDis},\ref{TDQTinTDDFT} are relegated to Appendixes I and II. In Sect. \ref{SecResults},
we report and discuss our results, for static and non-equilibrium regimes. 
Our conclusions are in Sect. \ref{Conclude}.
%
%
%

\section{The model}\label{sec_hamilt}
In standard notation, the lattice systems considered in this paper are described by the following Hamiltonian:
\begin{widetext}
\begin{eqnarray}
H=-\sum_{\sigma}  \sum_{l=-\infty}^{\infty} V_{l,l+1} (a^\dagger_{l\sigma}a_{l+1,\sigma}+H.c.)
+\sum_{\sigma} \sum_{l=1}^{L} \left[ w_l(\tau)+ \frac{U}{2} \hat{n}_{l-\sigma}\right] \hat{n}_{l\sigma}
+b_S(\tau)\sum_{l<1;\sigma}\hat{n}_{l\sigma}+
b_D(\tau)\sum_{l>L;\sigma}\hat{n}_{l\sigma}\label{generalH}.
\end{eqnarray}
\end{widetext}
Eq.(\ref{generalH}) describes a central chain of length $L$ (the lattice sites 
with $1\le l\le L$) connected to a left and a right 1D lead [sites with $l<1$ and $l>L$, respectively.  
The third and fourth term in Eq. (\ref{generalH}) represent the time-dependent bias in the leads
($\tau$ is the time variable), which is applied at time $\tau \geq 0$ [often, in the literature, the leads
are also referred to as the source (S) and drain (D), hence the subscripts $S,D$ in the bias terms in Eq.(\ref{generalH})].
For the contacted chain, the hopping term $V_{l,l+1}=V$ always, i.e. we employ transparent boundary conditions (hereafter, $V \equiv 1$ is taken as the energy unit). 
The Hamiltonian of the isolated chain is obtained from the general one by putting $V_{0,1}=V_{L,L+1}=0$ in in Eq.(\ref{generalH}), and retaining only the sites labeled by $1\leq l \leq L$.
Looking more closely to the chain part of the Hamiltonian, we have Hubbard-like interactions
(the term proportional to $U$; we set $U>0$) and time-dependent onsite energies $w_l(\tau)$, which
is convenient to separate into a static and time-dependent part: $w_{l}(\tau) = \epsilon_{l} + v_{l} (\tau)$. 
In the presence of disorder, the $\epsilon_{l} $:s are distributed according to some disorder probability
distribution. In this work, we use primarily the box disorder distribution, i.e.  $\epsilon_{l}\in[-W/2,W/2]$, but, we will sometimes consider binary
disorder, where $\epsilon_l=\pm W/2$. In both cases, $W$ fixes the strength of the disorder.
The chain Hamiltonian is a finite-size realization of the so-called Anderson-Hubbard model (AHM)\cite{AHM},
one of the most used models to study strongly correlated and disordered systems  \cite{reviewdisorderinteraction}.  
The AHM generalizes the standard Hubbard Hamiltonian \cite{Hubbard}
to inhomogeneous (and, in our case, possibly time-dependent) situations. That is, 
$\hat{\mathcal{V}}(\tau)\equiv  \sum_{l\sigma}  v_{l} (\tau)  \hat{n}_{l\sigma}$ 
describes a local (in space and time), time-dependent potential in the chain.
In the static case (i.e. before the systems starts to time-evolve), all $v_{l}(\tau)=0$.
Furthermore, the usual Hubbard model for the chain \cite{Hubbard} is recovered when $w_{l}(t)=0$,
whilst, when $U=0$ but $\epsilon_{l}\neq 0$, the chain is described the so-called Anderson model of disorder \cite{Anderson58}.

\section{(TD)DFT for lattice models} \label{genaspDFT}

\subsection{General aspects of lattice (TD)DFT}
A DFT based on the site occupation numbers $n_R$ was 
introduced more than two decades ago, to describe some ground state properties of the Hubbard model
\cite{GSDFT1,GSDFT1bis, GSDFT2}. An exact LDA (based on the Bethe-Ansatz) for the inhomogeneous 1D Hubbard model
was first considered in Ref. \onlinecite{GSN}. Further significant progress came
when an explicit and simple expression for the XC functional based on the Bethe-Ansatz 
was provided \cite{LimaPRL03}, and practically used to
investigate different inhomogeneous Hubbard-type models.
In subsequent work, The LDA based on the Bethe-Ansatz for $v_{xc}$ was scrutinized against exact results 
\cite{LimaPRL03, LimaEPL02, CapellePRB, CapellePolini, FrancaCapelle}, providing energies,
particle densities and entropies with an accuracy within a few percents.

Recently, lattice DFT has also been used to determine the polarizability of the 1D Hubbard model \cite{AkandeSanvito10}, and also to study the  entanglement entropy of the Hubbard model  \cite{FrancaCapelle}. Furthermore,
explicit analytical expressions for the XC potentials in small clusters can be found
in Ref. \onlinecite{CarrascalFerrer}, while the role of the temperature on $v_{xc}$ has been discussed in Ref. \onlinecite{StefanucciKurth}.  Lattice (TD)DFT has also been used to investigate ultracold atoms on an 1D optical lattice \cite{CapellePolini, Polini08, Hu2010,Xianlong2010, KarlssonEPL2011, DKAPCV11}. These systems permit to study different ground-state and non-equilibrium scenarios for the Hubbard model \cite{coldgeneral} with high accuracy (because a precise tunability of the lattice parameters is possible), more directly and easily than in solid-state experiments. 
In 2D, the Hubbard model has been investigated via DFT on the graphene lattice \cite{Harju}.
To date, the simple cubic lattice is the only 3D case considered in the literature \cite{DKAPCV11},
with the ground-state energy of the uniform system computed within dynamical mean field theory (DMFT)
\cite{metznervollhardt,revdmft}. Differently from the 1D case, here a discontinuity in $v_{xc}$ appears only for $U > U_c^{Mott}$, a DFT description of the onset of the Mott-Hubbard metal-insulator regime 
at a finite $U$.  

While lattice DFT rests on rigorous grounds \cite{GSN}, there is at present no direct formulation of the Runge-Gross theorem for lattice TDDFT,
as discussed in Ref. \onlinecite{KurthStefanucci2011}; early discussions of $v_0$-representability 
on the lattice can be found in Refs. \onlinecite{Baer08, Ullrich08, Verdozzi08}.
On the other hand, with the bond-current as the basic variable, 
a rigorous formulation on the lattice 
becomes possible \cite{GSEPMC10, Tokatly11, KurthStefanucci2011}.
Finally, it should also be noted that in 1D systems, as those considered in this paper, 
there is a one-to-one correspondence between densities and currents, and thus
TDDFT rests on solid grounds. 
Lattice (TD) DFT has also been considered in context of 
work on quantum transport geometries; this aspect will be examined further below.

\subsection{Formulation}
In this paper, we confine ourselves to the non magnetic 1D case; we review here the actual formulation for spin-independent (TD)DFT. 
In standard DFT notation, we can write for the ground-state total energy \cite{GSDFT2,GSN}:
\begin{align}
E[n,v_{ext}] \equiv T_0[n] + E_H[n] + E_{xc}[n]+\sum_i v_{ext}(i) n_i,
\label{theory::Exc}
\end{align}
\noindent where $v_{ext}(i) \equiv \epsilon_i$ is the static external field,
and $T_0[n]$ and $E_H = \frac{1}{4} \sum_i U_i n_i^2$ are, respectively, 
the non-interacting kinetic energy and the Hartree energy, with $n_i=\sum_\sigma n_{i\sigma}$.
To perform a local density approximation,  $E_{xc}$ is obtained from a homogeneous reference system (Hubbard model):
\be
E_{xc} = E - T_0- E_H \;.
\label{HUB_EXC}
\ee
To obtain $v_{xc}$, one takes the derivative of the XC energy per site $e_{xc}\equiv E_{xc}/L$
with respect to the density (in the general case,  
this should be a functional derivative):
\be
v_{xc}=\frac{\partial  e_{xc}(n,U)}{\partial n}\;.
\label{HUB_VXC}
\ee
For bipartite lattices, $e_{xc}(n,U)= e_{xc}(2-n,U)$ in the entire density range $[0,2]$ and thus  
$v_{xc}(n)=-v_{xc}(2-n)$. 
Finally, a local-density approximation is defined: 
\be
v_{xc}(i) = v_{xc} (n_i).
\ee
In ground-state DFT-LDA calculations, the XC potential obtained in this way is used to to solve self-consistently the Kohn-Sham (KS) equations
\begin{equation}
( \hat{t} + \hat{v}_{KS} ) \varphi _\kappa = \varepsilon_\kappa\varphi _\kappa \ \ ,    \label{staticKS}                             
\end{equation}
where $\hat{t}$ denotes the matrix for the single-particle hoppings  between nearest-neighboring sites, 
 and $\varphi_\kappa$
is the $\kappa$-th single-particle KS orbital, with $n_i = \sum_{\kappa \in occ} |\varphi _\kappa (i)|^2$.
The effective potential matrix is diagonal: $(\hat{v}_{KS})_{ii}=v_{KS} (i) = v_H(i) + v_{xc} (i) + v_{ext}(i)$, with  $v_H(i)=\frac{1}{2} U_i n_i$
being the Hartree potential.

If DFT can be a viable route to describe the ground-state properties
of Hubbard-type models, then the lattice Kohn-Sham (KS) equations could be
propagated in time, to get a TDDFT description
of the dynamics of lattice systems.
For 1D Hubbard-type Hamiltonians, work in this direction was performed in Refs. \onlinecite{Arya1, Arya2, Magyar}
for the linear response regime. 
A TDDFT approach to the real-time dynamics of the Hubbard model out of equilibrium 
was first considered in Ref. \onlinecite{Verdozzi08}, where exact results for the density and the XC potentials
were compared to those obtained by solving the time-dependent KS equations 
\be
\left( \hat{t} + \hat{v}_{KS}(\tau)\right ) \varphi_\kappa(\tau) = i \partial_\tau  \varphi _\kappa(\tau)\;,\label{TD_KS}                                
\ee
In general, $ v_{KS} (i,\tau) = v_H(i,\tau) + v_{xc} (i,\tau) + v_{ext}(i,\tau)$ depends non-locally 
on the density via $v_{xc}$. The adiabatic local density approximation (ALDA) \cite{Soven} to the XC potential
is then obtained with the prescription 
$v^{ALDA}_{xc}(i,\tau)\equiv v^{LDA}_{xc}(n_i(\tau))$, where 
the TD density is given by $ n_i(\tau) = \sum^{occ} _{\kappa} |\varphi _\kappa (i,\tau)|^2$. 
An ALDA for the Hubbard model was first introduced in Ref. \onlinecite{Verdozzi08}, with the treatment limited to spin-compensated systems, while the spin-dependent case was presented in Ref. \onlinecite{Polini08}, where TDDFT results and time-dependent DMRG
(tDMRG) results were compared.

For finite systems, a study more focussed on the role of non-local and memory effects 
beyond the ALDA was performed in Ref. \onlinecite{VKPAB_ChemPhys11}, via the
ALDA, exact, and Kadanoff-Baym time-evolution in small cubic
Hubbard clusters. The Kadanoff-Baym equations (KBE), with a many-body perturbation-theory approach 
to the self-energy, permit to take into account 
non-locality and memory effects on equal footing.
Such comparisons showed that an ALDA coming from the appropriate 
(strongly correlated) reference system can perform well in many instances,
(especially for slow perturbations) but, quite generally, it will  fail
for fast perturbations, or very strong interactions.

We conclude this Section with a remark about notation: the one adopted throughout the paper is fully consistent with the continuum case,
i.e. as if the Hubbard interaction was treated as spin-independent:
$\hat{U} = \frac{1}{2} \sum_i U_i(\hat{n}^2_i - \hat{n}_i)$. However, when the interaction is rewritten as in Eq. (\ref{generalH}) 
the interaction is effectively kept among opposite spins (i.e., treated as spin-dependent) \cite{equivalent} and the exchange has been removed at the Hamiltonian level. Thus,
$e_{xc}$ and $v_{xc}$ in Eqs. (\ref{HUB_EXC},\ref{HUB_VXC}) contain only 
correlation, and the Hartree plus exchange potential is $Un_i/2$. More aptly, our DFT quantities 
could have been called $e_c$ and $v_c$  but, following a common practice in the literature on lattice 
(TD)DFT, we still denote them by $e_{xc}$ and $v_{xc}$.
%
\subsection{Obtaining $v_{xc}$ for the 1D Hubbard model}\label{DFTHub}
According to Eq.(\ref{HUB_EXC}), to construct a LDA in 1D we need the exact ground-state energy of the infinite homogeneous 1D Hubbard model, where the hopping $V_{l,l+1}\equiv V$ and the interaction is present at all sites. This requires \cite{GSN,LimaPRL03,CapellePolini,AkandeSanvito10}
to solve the coupled Bethe-Ansatz equations for the charge and spin distribution functions [$\rho(x)$ and  $\sigma(x)$, respectively] \cite{LiebWu}. An ALDA is then easily obtained \cite{Verdozzi08,Polini08}, making $v_{xc}$ to become a function of the instantaneous local density. In the non-magnetic case considered here
(where the spin-up and spin-down densities are equal, i.e. $n_\uparrow=n_\downarrow=n/2$), the Bethe-Ansatz equations 
read
\begin{widetext}
\begin{eqnarray}
\rho(x)&=&\frac{1}{2\pi}+\frac{\cos x}{\pi}\int_{-\infty}^{+\infty}\frac{u/4}{(u/4)^{2}+(y-{\normalcolor \sin}x)^{2}}\sigma(y)dy  \label{BAeqs1}\\
\sigma(y)&=&\frac{1}{\pi}\int_{-Q}^{+Q}\frac{u/4}{(u/4)^{2}+(y-{\normalcolor \sin}x)^{2}}\rho(x)dx-\frac{1}{\pi}\int_{-\infty}^{+\infty}\frac{u/2}{(u/2)^{2}+(y-y^{\prime})}\sigma(y^{\prime})dy^{\prime}, \label{BAeqs2}
\end{eqnarray}
\end{widetext}
with $u\equiv U/V$. 
The functions $\rho(x)$ and  $\sigma(x)$ are related to the charge $n=n_\uparrow+n_\downarrow$ and spin-down $n_\downarrow$ densities via
\begin{eqnarray}
\int_{-Q}^{+Q}\rho(x)dx=n, \;\;  \int_{-\infty}^{+\infty}\sigma(y)dy=n_{\downarrow} \label{SCnorm}, 
\end{eqnarray}
from which the integration limit $Q$ is determined.
The ground state energy density for \textrm{$n_{\downarrow}=n_{\uparrow}$} is given by
\begin{eqnarray}
e(n_{\uparrow} = n_{\downarrow},U)=-2V\int_{-Q}^{+Q}\rho(x)\cos xdx \label{gsenergy}.
\end{eqnarray}
Eqs.(\ref{BAeqs1}-\ref{gsenergy})
are the prescription used in this work to determine $e_{xc}$.
The numerical solution of Eqs. (\ref{BAeqs1},\ref{BAeqs2}) was
obtained via a self-consistent procedure, with $Q$ adjusted at 
each iteration via the normalization condition in Eq.(\ref{SCnorm}).
Numerical integrations were performed with a 128-point Gauss-Legendre
quadrature  and, for each $U$, $e_{xc}(n,U)$ was 
obtained at the nodes of a uniform mesh for the density $n$. 
To obtain $v_{xc}$ at each node of the density mesh, we computed 
$\delta e_{xc}/\delta n$ with a 5-point numerical derivative.
To calculate $v_{xc}$ at off-node densities, a linear interpolation 
between the the closest nodes was employed. The $v_{xc}$
thus obtained is discontinuous at half-filling, as it should be for the homogeneous 1D Hubbard model; however, for a finite interacting system
contacted to non-interacting leads, the discontinuity of the exact $v_{xc}$ becomes
slightly smoothened (this was indicated in Ref. \onlinecite{KSKVG10},
using support from small Anderson clusters, and fully discussed
in Refs. \onlinecite{Burke,StefanucciKurth,EvSchmitt_broad,stafford}). According to these 
considerations, and also for numerical convenience, the XC potential 
was slightly smoothened in our actual calculations.

\section{The role of disorder}\label{SectDis}
Methodologically, the way we numerically deal with disorder effects in this paper is  
straightforward, since in most cases we limit our analysis to the arithmetic (configuration) average of specific quantities: 
the inverse participation ratio (IPR), the density and the current density. In most cases, the numerical configuration averages are performed over an incomplete set of configurations. Sects. \ref{IPRproposal}
and \ref{CPAsection} constitute an exception. In these sections, which deal with ground-state quantities, and for the case of binary disorder, we also perform 
complete numerical averaging over all the configurations, to provide benchmark results. For brevity, in the following,
complete numerical averaging will be referred to as 'exact averaging'.

\subsection{The inverse participation ratio}\label{IPRproposal}

A quantity often used as an indicator of localization in a system is 
the so-called inverse participation ratio  $\zeta $. The original definition \cite{IPRorigin} of $\zeta$, introduced for
non-interacting disordered systems, characterizes a given one-particle quantum state $\psi$ as follows:
\begin{eqnarray}
\zeta_0= \sum_i^M n_i^2  \label{IPR2}/(\sum_i^M n_i)^2,  \label{IPR1}
\end{eqnarray}
where $n_i=|\psi_i|^2$ is the density at site $i$ and the sums extend to all the sites $M$ in the system. 
For completely localized states (when $\psi\neq 0$ at only one site) we get $\zeta_0=1$, while
 $\zeta_0$ is smallest for delocalized states. To deal with interacting systems, suitable modifications
of Eq.(\ref{IPR2}) are, for example,
\begin{eqnarray}
\zeta_1&=&\sum_i^M \Delta n_i^2 /(\sum_i^M \Delta n_i)^2, \label{IPRSh}\\
\zeta_2(\omega)&=&\sum_i^M n_i^2(\omega) /[\sum_i^M n_i(\omega)]^2. \label{IPRWo}
\end{eqnarray}
The use of $\zeta_1$ is convenient when dealing with small systems
with discrete many-body levels \cite{Shepelyanski}. In this case, 
for $N$ particles, $\Delta n_i$ is the difference
between the ground-state densities with $N+1$ and $N$ particles, a clear
operational prescription for finite systems. 
Using $\zeta_2$ \cite{Wortis} amounts to consider the density of states 
as obtained from the one-particle propagator, since $n_i(\omega)=-
\Im G_{ii}(\omega)/\pi$. 
It should be noted that most investigations of the IPR are done numerically,
for finite systems. Using $\zeta_2$ requires introducing 
a finite artificial broadening $\gamma$, and employing a finite-size scaling analysis,
to assess the role of $\gamma$  \cite{VollhardtDOS,Wortis2,Johri}. 

In this work, we study finite disordered systems (short chains) contacted to semi-infinite 
homogenous leads. While the definition of IPR via Eq.(\ref{IPRWo})
is in principle suitable, for our lead-device-lead system we are faced with two issues:
i) a sum is implied over all the sites {M} in the system, including
the leads; this considerably increases the difficulty of the calculation ii) more fundamentally,
the IPR defined in this way can reflect the properties of the leads rather than the actual 
systems, since the lead-contribution can dominate the sums in Eq.(\ref{IPRWo}).
In view of this, we suggest the following possible alternative definition of the energy dependent IPR:
\begin{eqnarray}
\zeta_3&=&\sum_{i\in C} n_i^2(\omega) /[\sum_{i\in C} n_i(\omega)]^2 \label{IPRUS},
\end{eqnarray}
i.e. the sum is confined to the region of the device. This modified
definition of the IPR will be further analyzed and compared to the standard one in Sect. \ref{IPRjustify}.

\subsection{The Coherent Potential Approximation \label{CPAsection}}
Among the possible theoretical approaches to disorder, an important place is occupied by 
the Coherent-Potential Approximation \cite{CPAa,CPA0,CPA1} (CPA), which
introduces a simplified, approximate treatment of disorder averaging \cite{CPA0,CPA1,CPA3,CPA4}. A major appeal of CPA in its simplest formulation is
the pedagogical value, a relative analytical simplicity and ease of numerical implementation, together with the ability to 
give results for several quantities (e.g. ground state energies, transport properties, phase stability, photoemission) which are generally in broad agreement with experiment. 
Furthermore, the CPA becomes exact in the $D=\infty$ limit \cite{CPA5}; in
finite dimensions, it has been numerically tested against accurate numerical averaging
\cite{CPA5.1} and it has also been used in conjunction with many-body perturbation
theory (see e.g. Ref. \onlinecite{CPA6}). It has also been shown \cite{CPA7} that, when possible, numerical averaging based on small optimized supercells can give results considerably superior
to the CPA. Furthermore, it should be pointed out that the CPA shows significant limitations when describing quantities where
spatial correlations among different sites cannot be neglected.
A discussion of the properties of CPA (its limits of validity, extensions, 
applications, etc.) is outside the scope of this work, and here
we provide a short derivation which combines CPA and lattice DFT for
quantum transport geometries. To this end, we must slightly modify the standard 
treatment\cite{CPA1,CPA3,CPA4}, and adapt it
to the case of a finite disordered sample in the presence
of homogeneous semi-infinite contacts, and within a lattice DFT context.
We consider the case of diagonal disorder, and
specialize to a finite sample (chain of length $L$) of a random binary alloy, 
with species A and B and concentrations $c_A=N_A/L$ and $c_B=1-c_A$, respectively.
In the quantum transport geometry we study, the chains are connected 
to non-interacting leads, as in Sect.\ref{TDQTinTDDFT}.
For such chains, complete disorder averaging for a given concentration
requires $\binom{L}{N_A}$ configurations, and already for $L=14,15$
(as considered in this work), this number is rather large. In fact,
performing time-dependent quantum transport calculations based
on complete numerical averaging is computationally very demanding,
and one is bound to use a much reduced (and incomplete) 
numerical sampling. This latter strategy is the one mainly
adopted in the paper. To assess the scope of CPA,we limit ourselves
to the ground state, when no bias is applied. Numerical results relative to
this formulation are presented in Section \ref{resultsCPA}.

In matrix notation (in the site indexes), and in absence of disorder and magnetic effects, the retarded one-body Kohn-Sham propagator for a chain connected to a left and a right lead can be written as  
\begin{eqnarray}
\hat{g}_{KS}(\omega)=\frac{1}{\omega \hat{1}-\hat{H}_{KS}-\hat{\sigma}_L(\omega)-\hat{\sigma}_R(\omega)},\label{propKS}
\end{eqnarray}
where the matrix indexes of $g_{KS}$ label the sites of the chain, $\hat{\sigma}_{L (R)}$ 
is the self-energy operator from the left (right) lead \cite{selfexpress}, and $\hat{H}_{KS}=\hat{t}+\hat{v}_{KS}$,
accounts for the kinetic and potential Kohn-Sham (KS) operators.
In the presence of binary diagonal disorder $\hat{H} \rightarrow \hat{H}_{KS}+\hat{V}$, where
$\hat{V}=\sum_i \epsilon_i \hat{n}_i$. Here, $\hat{n}_i=\sum_\sigma \hat{n}_{i\sigma}$.
For $\epsilon_i$, the probability distribution is $P(\epsilon_i)=c_A \delta(\epsilon_i-\epsilon_A)+ (1-c_A) \delta(\epsilon_i-\epsilon_B)$. 
The CPA provides a prescription to determine the effect of $\hat{V}$. In an exact fashion,
we can equivalently write for the propagator $\langle G_{KS} \rangle$ averaged over all disorder configurations (the local dependence on $\omega$ is not shown):
\begin{eqnarray}
\langle \hat{G}_{KS} \rangle &=& \hat{g}_{KS}+\hat{g}_{KS}\;\; \hat{\Sigma} \;\; \langle \hat{G} _{KS}\rangle \label{EqSig}\\
\langle \hat{G}_{KS} \rangle &=& \hat{g}_{KS}+\hat{g}_{KS}\;\; \langle \hat{\mathcal{T}} \rangle \;\; \hat{g} _{KS} ,\label{EqT}
\end{eqnarray}
($\hat{\mathcal{T}}$ is the T-matrix of the potential.)
Inserting Eq.(\ref{EqT}) in Eq.(\ref{EqSig}), and after some simple manipulations, we get
\begin{eqnarray}
	\hat{\Sigma}= \langle  \hat{\mathcal{T}} \rangle \left[1+ \hat{g}_{KS} \langle   \hat{\mathcal{T}} \rangle\right]^{-1}.
\end{eqnarray}
In the CPA, the correlations among different scatterers are taken into account by assuming 
an effective medium for which the single site scattering is zero on average. To see how, 
we start with a specific disorder configuration $\hat V$, for which $\hat{G}_{KS}= \hat{g}_{KS}+\hat{g}_{KS} \hat{V} \hat{G}_{KS}$. Then, we subtract from both sides the quantity $\hat{\Sigma} \hat{G}_{KS}$ (with $\hat{\Sigma}$ yet to be specified).
This gives $(\hat{g}^{-1}_{KS}-\hat{\Sigma})=1+(\hat{V}-\hat{\Sigma})\hat{G}_{KS}$ and, if  we choose
$(\hat{g}^{-1}_{KS}-\hat{\Sigma})=\langle \hat{G}_{KS} \rangle^{-1}$, i.e. as in Eq.(\ref{EqSig}),
we finally get
\begin{eqnarray}
\hat{G}_{KS} &= &\langle \hat{G} _{KS}\rangle + \langle \hat{G} _{KS}\rangle (\hat{V}-\hat{\Sigma})\hat{G}_{KS}\nonumber\\
			   &=&\langle \hat{G} _{KS}\rangle + \langle \hat{G} _{KS}\rangle \;(\hat{\mathcal{T}} )\;
\langle \hat{G} _{KS} \rangle \label{Tav}
\end{eqnarray}
Performing the average of Eq.(\ref{Tav}) over different configurations , we note that it must
be $\langle \hat{\mathcal{T}} \rangle=0$, the key exact condition for the T-matrix. 

The CPA makes two assumptions: i) $\hat{\Sigma}$ is diagonal in the site-indexes, $\hat{\Sigma}^{CPA}_{ij}=\delta_{ij}\Sigma^{CPA}_i$, and so is
the perturbation $\hat{V}-\hat{\Sigma}$; ii) instead of $\langle  \hat{\mathcal{T}} \rangle=0$, one imposes
a simpler approximate constraint, i.e. that the average local T-matrix at the i-th site is zero: 
\begin{eqnarray}
\langle  \textswab{t}(i) \rangle = 0 =c_A \langle \textswab{t}_A(i) \rangle + c_B \langle \textswab{t}_B(i) \rangle,\label{averloct}
\end{eqnarray}
where
\begin{eqnarray}
\textswab{t}_{A (B)} (i) = \frac{\epsilon_{A(B)}-\Sigma^{CPA}_i}{1-\langle \hat{G}_{KS} \rangle_{ii} 
[\epsilon_{A(B)}-\Sigma^{CPA}_i]}.\label{localt}
\end{eqnarray}
In contrast to the usual treatments \cite{CPA0,CPA1,CPA3,CPA4},
here all quantities in Eq.(\ref{localt}) depend on the site-index, since
our system exhibits no disorder in the leads. Very recently,
and independently, a similar formulation has been provided in Ref. \onlinecite{Vedyayev}.

Inserting Eq.(\ref{localt}) in Eq.(\ref{averloct}), and performing simple algebra,
we arrive to an equation for $\Sigma^{CPA}_i$:
\begin{eqnarray}
\Sigma^{CPA}_{i}-\epsilon_A = \frac{(1-c_A)(\epsilon_B-\epsilon_A)}{1-\langle \hat{G}_{KS} \rangle_{ii} 
[\epsilon_{B}-\Sigma^{CPA}_i]}.\label{localt1}
\end{eqnarray}
This equation must be solved for each site in the chain, once 
the local propagator $\langle \hat{G}_{KS} \rangle_{ii}$ is known. 
The latter is in turn
determined from
Eq.(\ref{propKS}), after the replacement $\hat{H}_{KS}\rightarrow \hat{H}_{KS}+\hat{\Sigma}^{CPA}$ is made, and after the dependence of $\hat{H}_{KS}$ on the densities
has been taken into account via
\begin{eqnarray}
n_i=-\int_{-\infty}^\mu \Im \langle \hat{G}_{KS}(\omega+i0^+)  \rangle_{ii}\frac{d\omega}{\pi}, \label{chem}
\end{eqnarray}
with $\mu$ the chemical potential (here, as in the rest of the paper, we work at zero temperature). While it is certainly possible (and often necessary) to improve over the CPA \cite{CPAreview}, in this work we aim at qualitative insight, and
in Sect \ref{resultsCPA} we present results obtained with the simple local formulation of CPA
and the self-consistent set of equations Eqs.(\ref{propKS},\ref{localt},\ref{chem}) \cite{howgroundstate}.
\section{Time Dependent Quantum Transport (TDQT) and TDDFT} \label{TDQTinTDDFT}

Theoretical approaches to quantum transport can be broadly grouped according
to different criteria, e.g. if they are based on a steady-state or time-dependent 
formulations, if they use {\it ab initio} or model
Hamiltonian methods, or according to which mathematical technique is employed:
non-equilibrium-propagator, linear-response, wavefunction-scattering, etc.
Here we consider time dependent quantum transport (TDQT), 
which permits to follow the system during its time evolution after a bias has been applied. 
In this way, steady-state, transient and a.c. currents can all be considered on equal 
footing and, in the presence of dissipation, history dependence (memory effects) 
are also accounted for. A viable strategy to TDQT is to consider large but
finite systems  \cite{DiVentrafinite}. Via an initial spatial imbalance of particles, 
a quasi-steady state current can be established. Recently this approach has also been
used to describe bosonic and fermionic transport of ultra-cold atoms in 1D optical lattices
\cite {diventralastcoldtransport,diventralastcoldtransport1}.

A different formulation, the one used here, 
considers a central region initially connected to semi-infinite leads \cite{Cini80,gstcoa,Petri2008}. 
This "contacted" approach has been used to introduce a TDDFT description of 
TDQT \cite{gstcoa,gstcoaEPL}, and the practical applicability of a TDDFT scheme
has also been shown \cite{practical}. Furthermore, classical nuclear degrees of freedom have
also been included in the approach\cite{cvgstcoa}.

In a TDDFT approach to TDQT \cite{gstcoa,gstcoaEPL}, a key quantity is the 
XC potential.  In Ref. \onlinecite{SchmittEver08},  a
combination of DMRG and lattice DFT was used to gain insight into the exact ground-state XC functionals for a correlated-electron model system coupled to external reservoirs. 
A comparison of lattice DFT and DMRG in transport has also been provided in Ref. \onlinecite{schenk2011}, whilst a study
of the role of spin in the XC potential can be found in Ref. \onlinecite{Thijssen} (for treatments based on 
ground-state current DFT for lattice models, see  Refs. \onlinecite{Schenk2, AkandeSanvito11}).

The effect of a discontinuity in $v_{xc}$ in a
TDDFT description of TDQT was examined within lattice TDDFT in the ALDA approximation \cite{KSKVG10}. Following the time-evolution of a single Anderson impurity attached to two biased leads, a dynamical notion of the Coulomb blockade was then presented. This emerges also from a description based on time-dependent, bond-current DFT\cite{KurthStefanucci2011}. In Ref. \onlinecite{KSKVG10}, it was also pointed out that for a single Anderson impurity, the exact $v_{xc}$ is a sharp (at half-filling ) but smooth function of the density. 

Subsequently, a comparison between ALDA, tDMRG and KBE for TDQT in lattice systems was presented in Ref.  \onlinecite{KBEALDAtransport11}, showing that the ALDA can give accurate densities but overestimated currents, due to the neglect
of non-local effects in the leads. We finally mention that, very recently, different research groups \cite{Burke,StefanucciKurth,EvSchmitt} independently pointed out that suitable XC potentials permit a (TD)DFT description of the Kondo effect, and also examined in detail the broadening of the 
derivative discontinuity of the XC potential \cite{StefanucciKurth,Burke,EvSchmitt_broad,stafford}.

\subsection{Time Evolution for Quantum Transport} \label{TEQT}
The time-dependent scheme used in this work is the one developed in Ref. \onlinecite{practical} and, as
in Ref. \onlinecite{KSKVG10}, interactions in the central region
are treated via an ALDA from the Bethe Ansatz \cite{Verdozzi08}. 
For disordered systems, where large central regions and
configuration averages may be needed, such an algorithm may be computationally expensive. As described in
Sect.\ref{Lanczadapt}, a convenient way  to enhance its numerical efficiency is to 
use the Lanczos recursion for 
time evolution \cite{JChemphys} (for a quick introduction to the Lanczos technique, see Appendix A).

We start with a concise description of the original algorithm \cite{practical},
as background to our Lanczos-adapted scheme, and we specialize to 1D geometries.
The notation in this and the next section
is closer to the one in Ref. \onlinecite{practical}, and thus differs from that 
in the rest of our paper.

The Hamiltonian we consider is  $\bH^{tot}(t)=\bH_{el}+\bU(t)$, where $\bU(t)$ is the external perturbation. 
In a TDDFT approach, the initial, ground state is a single Slater determinant $|\Q_{g}\ket$. It is useful to divide the (1D) space into three regions. With $s$ a the site label, we have the region $L$ (corresponding to the left lead,  with $s\le-(M+2)$ ), the central region C ( with  $|s|\le M+1$ , i.e the device region contains $2M+3$ sites), and the region $R$ (corresponding to right lead,  with  $s\ge(M+2) )$.  The general structure of any bound, extended or resonant one particle eigenstate $\q$ in the
Slater determinant $|\Q_{g}\ket$ can be written as 
\be
\q(s)
=\left\{
\begin{array}{ll}
L_{+}e^{-ik_{l} s}+L_{-}e^{ik_{l} s} & \;\; s\leq-M-2 \\
\q(s)  & \;\; |s| \leq M+1 \\
R_{+}e^{ik_{r } s}+R_{-}e^{-ik_{r} s} & \;\; s\geq M+2 \\
\end{array}
\right.,
\label{gsops}
\ee
To describe quantum transport, one needs to evolve in time the 
ground state configuration $|\Q_{g}\ket$, i.e. each one of the single particle eigenstates $\q$ above. 
Introducing the projection operators ${\bf P}_{L,C.R}$ (for example,  ${\bf P}_{L}=\sum_{s\in L}|s\ket\bra s|$), we can write ($\b=L,C,R$), for the generic single particle state,
\begin{equation}
|\q\ket=\sum_{\b}|\q_{\b}\ket,\quad |\q_{\b}\ket={\bf P}_{\b}|\q\ket.\label{short2}
\end{equation}
In the same way, we can project the Hamiltonian in the different regions
\begin{equation}
\bH=\sum_{\b\b'}\bH_{\b\b'},\quad \bH_{\b\b'}\equiv 
{\bf P}_{\b}\bH{\bf P}_{\b'}.\label{short1}
\end{equation}
Separating the contribution from the leads in $\bU$, the set of one-particle equations becomes
\be
i\frac{\rd}{\rd t}|\q(t)\ket=
\left[\bH(t)+\bU_{leads}(t)\right]|\q(t)\ket,
\ee
with $\bH(t)=\bH_{\rm el}+\bU_{CC}(t)$,
where  $\bH_{\rm el}$ 
is the electron one particle Hamiltonian and $\bU_{CC}(t)$ is the external potential projected in the central 
region $C$.
Assuming metallic electrodes,
\be
\bU_{leads}(t)
=\left\{
\begin{array}{ll}
\d_{s,s'}W_{L}(t) & \;\; s \leq -M-2 \\
0 & \;\; s \leq |M+1| \\
\d_{s,s'}W_{R}(t) & \;\; s \geq M+2 \\
\end{array}
\right..
\ee
In the numerical time propagation, the time is discretized:  $t_{m}=2m \d$, where 
$\d$ is the timestep, $m$ is an integer, and the explicit prefactor $2$ is introduced
for convenience in the formulas. In  \cite{practical}, the one-particle eigenstates are propagated from $t_{m}$ to $t_{m+1}$ using a generalized Crank-Nicholson scheme. 
For the time evolution of each one of the one-particle states in 
$|\Q_{g}\ket$, one gets \cite{practical}
\begin{eqnarray}
({\bf 1}+i\d \bH^{(m)})\frac{{\bf 1}+i\frac{\d}{2}\bU_{ 
leads}^{(m)}}{{\bf 1}-i\frac{\d}{2}\bU_{leads}^{(m)}}|\q^{(m+1)}\ket=\nonumber\\
({\bf 1}-i\d \bH^{(m)})\frac{{\bf 1}-i\frac{\d}{2}\bU_{ 
leads}^{(m)}}{{\bf 1}+i\frac{\d}{2}\bU_{leads}^{(m)}}|\q^{(m)}\ket,\;\;\;
\label{cn1}
\end{eqnarray}
where $|\q^{m}\rangle\equiv|\q(t_m)\rangle$ and
\begin{eqnarray}
\bH^{(m)}&=&\bH_{\rm el}+\frac{1}{2}\left[\bU_{CC}(t_{m+1})+\bU_{CC}(t_{m})\right]\\
\bU_{leads}^{(m)}&=&\frac{1}{2}[\bU_{leads}(t_{m+1})+\bU_{leads}(t_{m})].
\end{eqnarray}
Using Eqs.(\ref{short2},\ref{short1}), and after some algebra, the closed equation for the time-evolution in the central region is
\beq
|\q_{C}^{(n+1)}\ket=
\frac{{\bf 1}_{C}-i\d \bH^{(n)}_{\rm eff}}
{{\bf 1}_{C}+i\d \bH^{(n)}_{\rm eff}}|\q_{C}^{(n)}\ket\nonumber\\
-2i\d\sum_{\a=L,R}\frac{\Omega_{\a}^{(n)}}{w_{\a}^{(n)}}
\left(
|\s_{\a}^{(n)}\ket+|\m_{\a}^{(n)}\ket
\right),
\label{central}
\eeq
where
\begin{eqnarray}
w_{\a}^{(n)}=\frac{1-i\frac{\d}{2}W_{\a}^{(n)}}{1+i\frac{\d}{2}W_{\a}^{(n)}},\\
\Omega_{\a}^{(n)}=\prod_{j=0}^{n}[w_{\a}^{(j)}]^{2},
\end{eqnarray}
and
\begin{eqnarray}
\bH^{(n)}_{\rm 
eff}=\bH^{(n)}_{CC}-i\d\sum_{\a=L,R}\bH_{C\a}\frac{1}{{\bf 1}_{\a}+i\d 
\bH_{\a\a}}\bH_{\a C}\nonumber\\
=\bH^{(n)}_{CC}-i\d\sum_{\a=L,R}\bQ_{\a}^{(0)}.\label{Hcentral}
\end{eqnarray}
The $\bQ_{\a}^{(0)}$ matrices have only one non-zero element,
\be
\left[\bQ_{\a}^{(0)}\right]_{s,s'}=b^{(0)}
\left\{
\begin{array}{ll}
\d_{s,-M-1}\d_{s',-M-1} & \quad \a=L \\
\d_{s,M+1}\d_{s,M+1} & \quad \a=R
\end{array}
\right.,
\label{qmm}
\ee
with $b^{(0)}=\frac{-1+\sqrt{1+4\d^{2}V^{2}}}{2\d^{2}}$ and $V$ the hopping parameter
in the leads. 
The expression for the source state $|\s_{\a}^{(n)}\ket$ and the memory 
state
$|\m_{\a}^{(n)}\ket$ are \cite{practical}:
\begin{eqnarray}
|\m_{\a}^{(n)}\ket&=&Z_{\a}^{(n)}
\frac{1}{{\bf 1}_{C}+i\d \bH^{(n)}_{\rm eff}}|u_{\a}\ket,\\
|\s_{\a}^{(n)}\ket&=&G_{\a}^{(n)}
\frac{1}{{\bf 1}_{C}+i\d \bH^{(n)}_{\rm eff}}|u_{\a}\ket
\label{efss}
\end{eqnarray}
where $|u_{\a}\ket$ is a unit vector such that
\be
\bra s|u_{\a}\ket=\left\{
\begin{array}{ll}
\d_{s,-M-1} & \quad \a=L \\
\d_{s,M+1} & \quad \a=R 
\end{array}
\right..
\ee
The scalar quantities $Z_{\a}^{(n)}$ and $G_{\a}^{(n)}$, $\a=L,R$ are given by \begin{widetext}
\begin{eqnarray}
Z_{\a}^{(n)}&=&\frac{\d}{2i}
\sum_{j=0}^{n-1}
\frac{w_{\a}^{(j)}}{\Omega_{\a}^{(j)}}
\left(b^{(n-j)}+b^{(n-j-1)}\right)
\left(\bra u_{\a}|\q_{C}^{(j+1)}\ket+\bra u_{\a}|\q_{C}^{(j)}\ket\right),\label{open1}\\
G_{\a}^{(n)}&=&
\left(
\a_{+}e^{iz_{\a}(M+2)}+\a_{-}e^{-iz_{\a}(M+2)}
\right)
V\frac{\left(1-2i\d\cos\left(z_{\a}\right)\right)^{n}}
{\left(1-2i\d\cos\left(z_{\a}\right)\right)^{n+1}}
\label{sst} \\ 
&+&
\left(
\a_{+}e^{iz_{\a}(M+1)}+\a_{-}e^{-iz_{\a}(M+1)}
\right)
\times
i\d\sum_{j=0}^{n}
\frac{\left(1-2i\d\cos\left(z_{\a}\right)\right)^{n-j}}
{\left(1-2i\d\cos\left(z_{\a}\right)\right)^{n+1-j}}
\left(b^{(j)}+b^{(j+1)}\right)\nonumber
\end{eqnarray}
\end{widetext}
and $z_{\a}=k_l$ for $\a=L$ while $z_{\a}=k_r$ for $\a=R$.
For $n\ge2$, the quantities $b^{(n)}$ in the Eqs.(\ref{open1},\ref{sst})
are obtained by recursion:
\beq
b^{(n)}&=&
\frac{b^{(1)}b^{(n-1)}}{b^{(0)}}-\d^{2}\frac{b^{(0)}b^{(n-2)}}{1+2\d^{2}b^{(0)}}
\label{qm} \\ 
&-&\d^{2}
\sum_{j=1}^{n-1}\frac{\left(
b^{(j)}+b^{(j-1)}+b^{(j-2)}\right)
b^{(n-2-j)}}{1+2\d^{2}b^{(0)}}
\nonumber
\eeq
and  
$b^{(n<0)}=0, b^{(1)}=\frac{1-2\d^{2}b^{(0)}}{1+2\d^{2}b^{(0)}}b^{(0)}$ and $b^{(0)}$ the same as in
Eq.(\ref{qmm}).
\subsection{Lanczos-adapted algorithm}\label{Lanczadapt}
The basic idea behind the algorithm discussed in the previous Section
is to discretize the time axis via the Crank-Nicholson algorithm {\em before} performing
the partitioning in L, C, R regions \cite{practical}. One could think of doing the same
using the Lanczos algorithm for the time propagation \cite{JChemphys} (the method is quickly reviewed in Appendix A); however, non-commuting parts of the Hamiltonian would appear in
the exponent this time, rendering formal manipulations more involved. Here, we consider a simple shortcut that, while improving the numerical efficiency of the algorithm of Sect. \ref{TEQT}, has the same
degree of accuracy ( i.e. it is second-order in $\delta$) but avoids working with the
Lanczos scheme before the partitioning \cite{shortversion}.
Looking at Eq.(\ref{central}), we notice that the explicit action of $\bH^{(n)}_{\rm eff}$
occurs in two specific terms:
\begin{eqnarray}
|\chi_1\rangle&=&\frac{{\bf 1}_{C}-i\d \bH^{(n)}_{\rm eff}}
{{\bf 1}_{C}+i\d \bH^{(n)}_{\rm eff}}|\q_{C}^{(n)}\ket\label{numb1}\\
|\chi_2\rangle&=&\frac{1}{{\bf 1}_{C}+i\d \bH^{(n)}_{\rm eff}}|u_{\a}\label{numb2}\ket
\end{eqnarray}
where $|\chi_1\rangle$ is the contribution to $|\q_C^{(n+1)}\rangle$ from the central region, and  $|\chi_2\rangle$ enters the expressions for the source and memory states.
For $|\chi_1\rangle$, since $\delta\rightarrow 0$, one can write, up to order two in $\delta$
\be
|\chi_1\rangle=\frac{{\bf 1}_{C}-i\d \bH^{(n)}_{\rm eff}}
{{\bf 1}_{C}+i\d \bH^{(n)}_{\rm eff}}|\q_{C}^{(n)}\ket\approx e^{-2i\delta \bH^{(n)}_{\rm eff}}|\q_{C}^{(n)}\ket.
\ee
For the case of $|\chi_2\rangle$, we define the following quantities:
\beq
\Delta_\pm=\frac{1\pm \sqrt{3}}{2}\delta,
\eeq
which permit to rewrite $|\chi_2\rangle$ as
\begin{eqnarray}
\!\!\!\!\!\!\!\!|\chi_2\rangle=\left[-1+e^{-i\bH^{(n)}_{\rm eff}\Delta _+}
+e^{-i\bH^{(n)}_{\rm eff}\Delta _-}\right] |u_{\a}\ket +O(\delta^3)\label{nice}
\end{eqnarray}
If necessary, one can go to higher orders, by imposing that $(1+\delta x)^{-1}=A+\sum_k e^{a_k \delta x}$
and finding the coefficients $A, \{a_k\}$ by comparison of the two expressions order by order in $\delta$
(in general, the $ \{a_k\}$ will be complex). We note that the same Lanczos sequence of 
basis vectors is required for both exponentials in Eq.(\ref{nice}).\newline
All terms which appear in the propagation scheme of Sect. \ref{TEQT} and that involve  ${\bH^{(n)}_{\rm eff}}$, have been re-expressed in terms of exponentials, so that Lanczos propagation can be
used; finally, since ${\bH^{(n)}_{\rm eff}}$ is complex, Eq.(\ref{Hcentral}), it is convenient to split the exponentials;  for small
$\delta$, 
\begin{eqnarray}
e^{-2i\delta \bH^{(n)}_{\rm eff}}&\approx& e^{-\delta^2\sum_\alpha\fff^{(0)}_{\alpha}}
e^{-2i\delta \bH^{(n)}_{CC}}
e^{-\delta^2\sum_\alpha \fff^{(0)}_{\alpha}}\label{split}\\
e^{-i\Delta_\pm \bH^{(n)}_{\rm eff}}&\approx&
e^{-\frac{\delta}{2} \Delta_\pm \sum_\alpha \fff^{(0)}_{\alpha}}
e^{-i\Delta_\pm  \bH^{(n)}_{CC}}
e^{- \frac{ \delta}{2} \Delta_\pm \sum_\alpha \fff^{(0)}_{\alpha}}\nonumber\\\label{split2nd}
\end{eqnarray}
For the 1D case, the advantage is immediate:  the $\bQ_{\a}^{(0)}$ in Eq.(\ref{qmm})
have only one non-vanishing entry and
the outer exponentials in Eqs.(\ref{split},\ref{split2nd}) reduce to scalars. We expect that the splitting 
will still provide a simplification in the 3D case. To summarize, on increasing the size of the central region,
our Lanczos adapted scheme becomes highly convenient, potentially a significant advantage when
dealing with disordered and large samples.

%
%
%
%
\section{Results}\label{SecResults}
Recently, a non-perturbative study of finite Anderson-Hubbard chains
has been performed in terms of Density Matrix Renormalization Group and a real-space 
version of Dynamical Mean Field Theory \cite{juliawern}.
Using different indicators of delocalization, such as the geometrically
averaged LDOS and the IPR, the main outcome of such non-perturbative calculations 
was a clear indication of a tendency to delocalization in a range of $U, W$ values
in the ground state of these chains. Can a similar behavior be observed
in shorter chains contacted to semi-infinite leads? To address this issue,
we present here ground-state and dynamical results for short, isolated/contacted disordered and interacting chains.
No spin-effects are considered, i.e. the systems are spin compensated, and
all applied potentials are spin-independent.
The quantities we will analyze are the on-site particle density, the bond current, and the IPR. 
In general, we consider box-disorder, with onsite energies $\epsilon_l$ uniformly distributed in the interval $[-W/2,W/2]$. In this case, configuration averages will be done over a finite number of configurations. 
However, we used binary disorder to discuss the IPR (Sec. \ref{IPRjustify}) and the CPA (Sec. \ref{resultsCPA}), since, for short chains, exact averaging can be done with a manageable number of configurations.
Further details of each set/type of calculations will be provided in the respective sections. 
\subsection{Changing the definition of the IPR}\label{IPRjustify}
To analyze our definition of IPR, Eq.(\ref{IPRUS}), we find it convenient to consider binary (rather than uniform) disorder in
a non-interacting chain of $L=14$ sites. For binary disorder, choosing $L$ small and even
permits i) to consider exactly the $A_{50}B_{50}$ alloy concentration and ii) to perform disorder averages exactly.
The chain can be isolated or contacted to two 1D leads
(one at each end of the chain); the leads can be finite or semi-infinite (their length
will be denoted by $N_{ld}$). The total number of sites in the system is thus $M=L+2N_{ld}$.
\begin{figure}
\includegraphics[width=8.7cm]{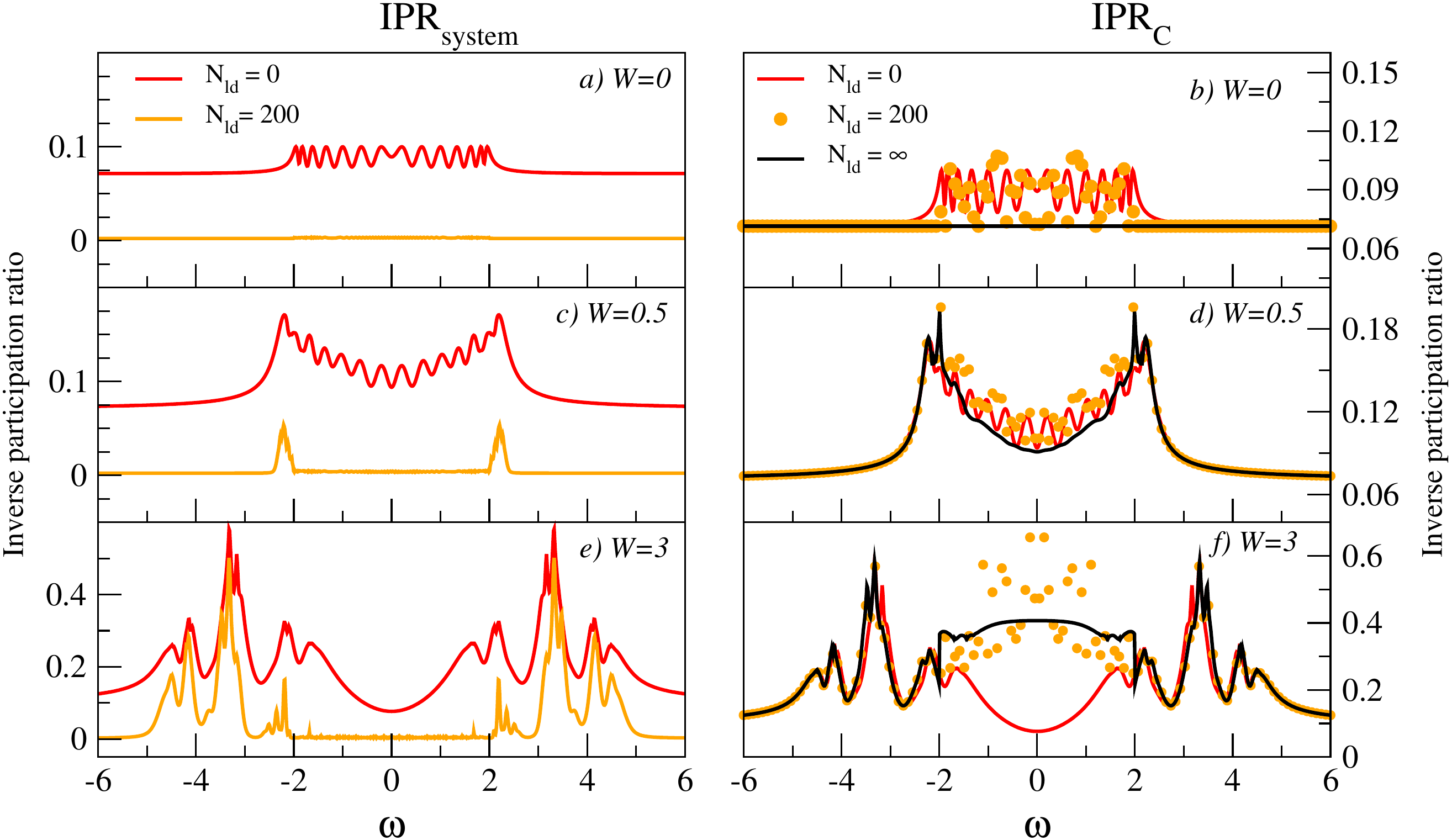}
\caption{\label{IPRfig1} (Color online) Disorder averaged, energy dependent IPR (see main text for the definition) for a $A_{50}B_{50}$, non-interacting binary-alloy chain C with $L=14$ sites, 
contacted to in/finite leads. Complete disorder averages were performed over $\binom{L}{N_A}=3432$ configurations. Top-to-bottom panels show cases with increasing site-disorder $W$. Left (right) panels show results
for an IPR estimated on the entire system (only on the chain C), as in Eq.(\ref{IPRWo}) [Eq.(\ref{IPRUS})]. The color/symbol coding in panel a) [b)],
also applies to panels c),e) [ d),f)]. In the left panels, $IPR_{system}(\omega)$ for infinite leads ($N_{ld}\rightarrow \infty $) is not shown. All curves were
obtained with a small Lorentzian broadening $\gamma$, to
have a minimum accuracy of $10^{-4}$ for $IPR_C(\omega)$ when $N_{ld}\rightarrow \infty $ [black curves in panels b),d),f)]. 
}
\end{figure}

We wish at this point to make a short technical digression on the numerical calculation of the IPR. 
For the energy-dependent IPR, we need the local density of states (LDOS) at site(s) $i$:
\begin{eqnarray}
n_i(\omega)=\frac{\gamma}{\pi}\sum_\lambda \frac{|\langle \lambda|i\rangle|^2}{(\omega-E_\lambda)^2 + \gamma^2}, \label{LDOS}
\end{eqnarray}
where $\lambda$ labels the one-particle eigenstates $|\lambda\rangle$ and eigenvalues $E_\lambda$ of the system. 
For infinite systems, Eq.(\ref{LDOS})  is not usable directly, and one resorts to Green's functions [see Eq.(\ref{propKS}) in Sect. \ref{CPAsection}].
For finite systems, one can e.g. use recursion techniques  \cite{Wortis2}, or, as done in this Section, perform a direct diagonalization of the disordered Hamiltonian. However, if the IPR must be determined in a range of energies (i.e. for several $\omega$ values), already for moderate system sizes $M$,
the $\lambda$- and $i$-sums in Eqs.(\ref{IPRWo},\ref{IPRUS},\ref{LDOS}) become computationally expensive. In Appendix B, we present a technique which permits to perform such nested summations in a rather efficient way. 

Results for the IPR according to the two definitions $\zeta_2$, Eq.(\ref{IPRWo}), and $\zeta_3$, Eq.(\ref{IPRUS}), are shown in Fig.\ref{IPRfig1}
 (hereafter, $\zeta_2$ and $\zeta_3$ will be renamed $IPR_{system}$ and  $IPR_{C}$, 
respectively).
Calculations with $IPR_{system}$ are reported in panels a,c,e). 
In all three panels, we see that on increasing the size $N_{ld}$ of the (finite) leads, $IPR_{system}(\omega)$ gets quickly reduced
in the region $|\omega|\leq 2$,  i.e. for the energy range for the extended states in the system (results for larger
$N_{ld}$, not shown, confirm this trend). 
Outside the band region, the decay of $IPR_{system}(\omega)$ on increasing $N_{ld}$ is much slower, 
and our numerical evidence, together with arguments based on the large $W$ limit,
shows that for larger $N_{ld}$   $IPR_{system}(\omega)$ vanishes everywhere for $|\omega|\ge 2$ except at the energies of the localized states, where it takes the corresponding  IPR value.
Thus, irrespective of the disorder strength in the finite chain, in the limit of semi-infinite leads, $IPR_{system}$ indicates
delocalization for $|\omega|\leq 2$. However, for a large disorder, the chain becomes disconnected from the leads, and this
is missed in the vanishing $IPR_{system}$, which simply reflect almost everywhere the 
delocalized states in the disconnected leads (as in most quantum transport treatments, the leads 
are assumed to be homogeneous and non-interacting). 

In the panels b,d,f) of Fig.\ref{IPRfig1}, we show results for $IPR_{C}(\omega)$; in this case,
on increasing $N_{ld}$, the IPR tends to a finite value (the asymptotic value for when $N_{ld}\rightarrow \infty$), 
which better reflects the fact that the localization
in some region of the system affects the system as a whole (we have also verified that on increasing
$W$, $IPR_{C}$ increases). Being a local quantity, $IPR_{C}$ obviously depends on the size and the details of the chain.

When interactions among particles are taken into account, there is another point that is necessary to examine. This aspect
is specific to our approach to quantum transport, where interactions are described within lattice (TD)DFT. 
Since the Kohn-Sham system is just a fictitious system apt to reproduce the true interacting density, a frequency dependent IPR of the Kohn-Sham system
has actually little physical meaning. Thus, our final proposed definition of IPR is:
\begin{eqnarray}
IPR^{KS}_C=\sum_{i\in C} [n_i^{KS}]^2 / [\sum_{i\in C} n_i^{KS}]^2 \label{IPRUS2}.
\end{eqnarray}
Eq.(\ref{IPRUS2}), which makes use of the actual particle density, can also be
used in the interacting and time-dependent cases, and thus is both conceptually and operationally well defined within (TD)DFT.  We have, in the same way as for the energy dependent IPR, verified that also this definition is sensitive to localization and tends to a finite value when the size of the leads tend to infinity.
To summarize, while not arguing that our definition of IPR is an optimal or unique indicator of localization in quantum transport geometries \cite{LucaMol97}, in our simulations we used Eq.(\ref{IPRUS2}) as a viable 
prescription. 
\subsection{CPA-DFT for short chains attached to leads}\label{resultsCPA}
\begin{figure*}
\includegraphics[width=13.cm]{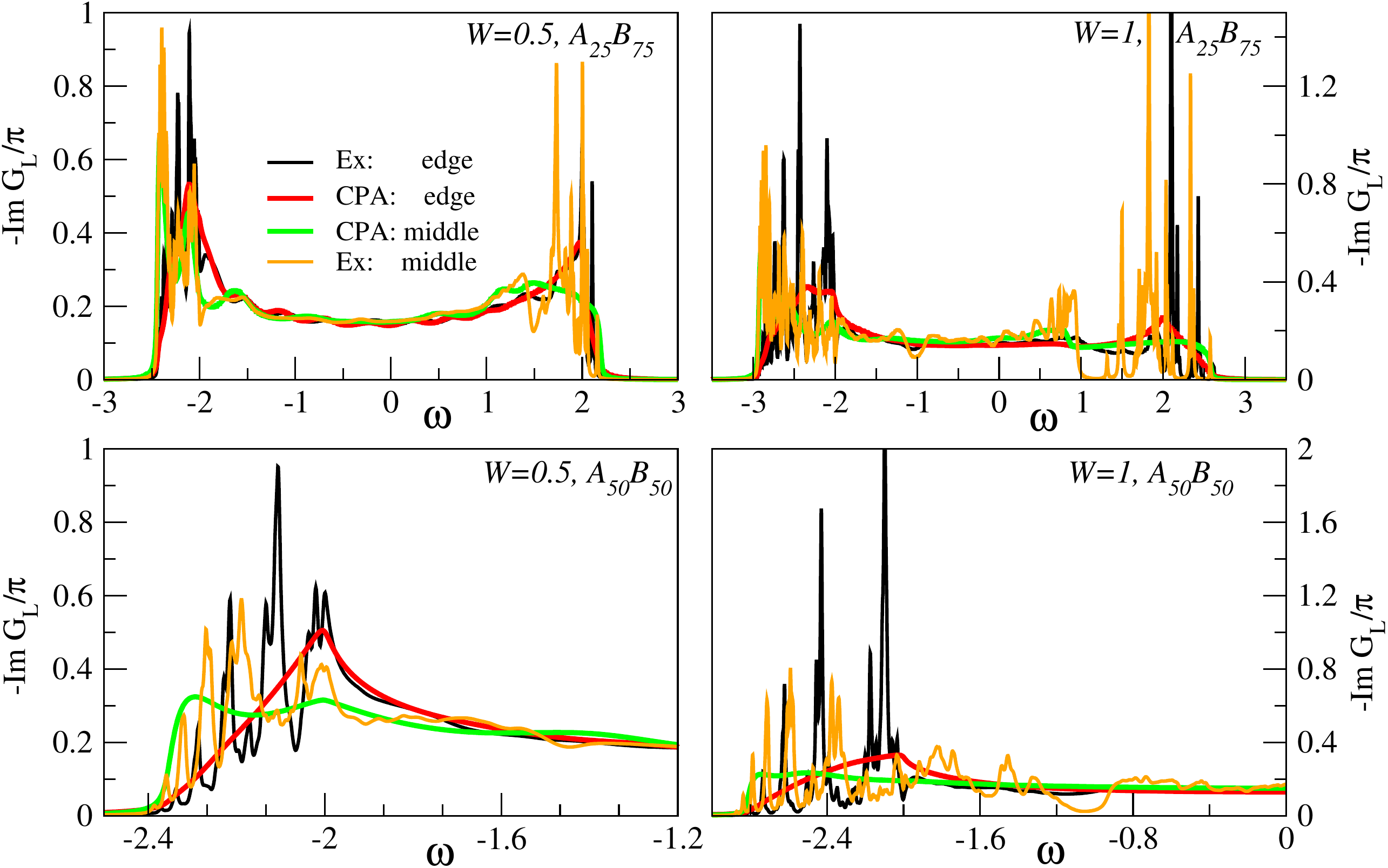}
\caption{\label{CPAfig1} (Color online) CPA versus exact averaging for a chain with 14 sites connected to two semi-infinite homogenous leads,
with transparent boundary conditions. The meaning of each curve is given by the color coding in the top-left panel. In the legend, {\it Ex.Av.} means complete numerical averaging.
The density of states for the leads is nonzero in the interval $[-2,2]$ and the chemical potential
$\mu=0$ in the leads (half-filling). In each panel, the average LDOS is presented for middle and edge sites (see main text), and a obtained with CPA (red and green curves) and exactly (black and orange curves), and two strengths of disorder $W$ are considered. For the $A_{50}B_{50}$ systems (bottom panels) all curves are symmetric w.r.t. $\omega=0$. For $W=1$, the maxima of some of the sharp peaks at the band-edges are outside the scale
shown in the figure.
}
\end{figure*}
\begin{figure}[b]
\includegraphics[width=8.5cm]{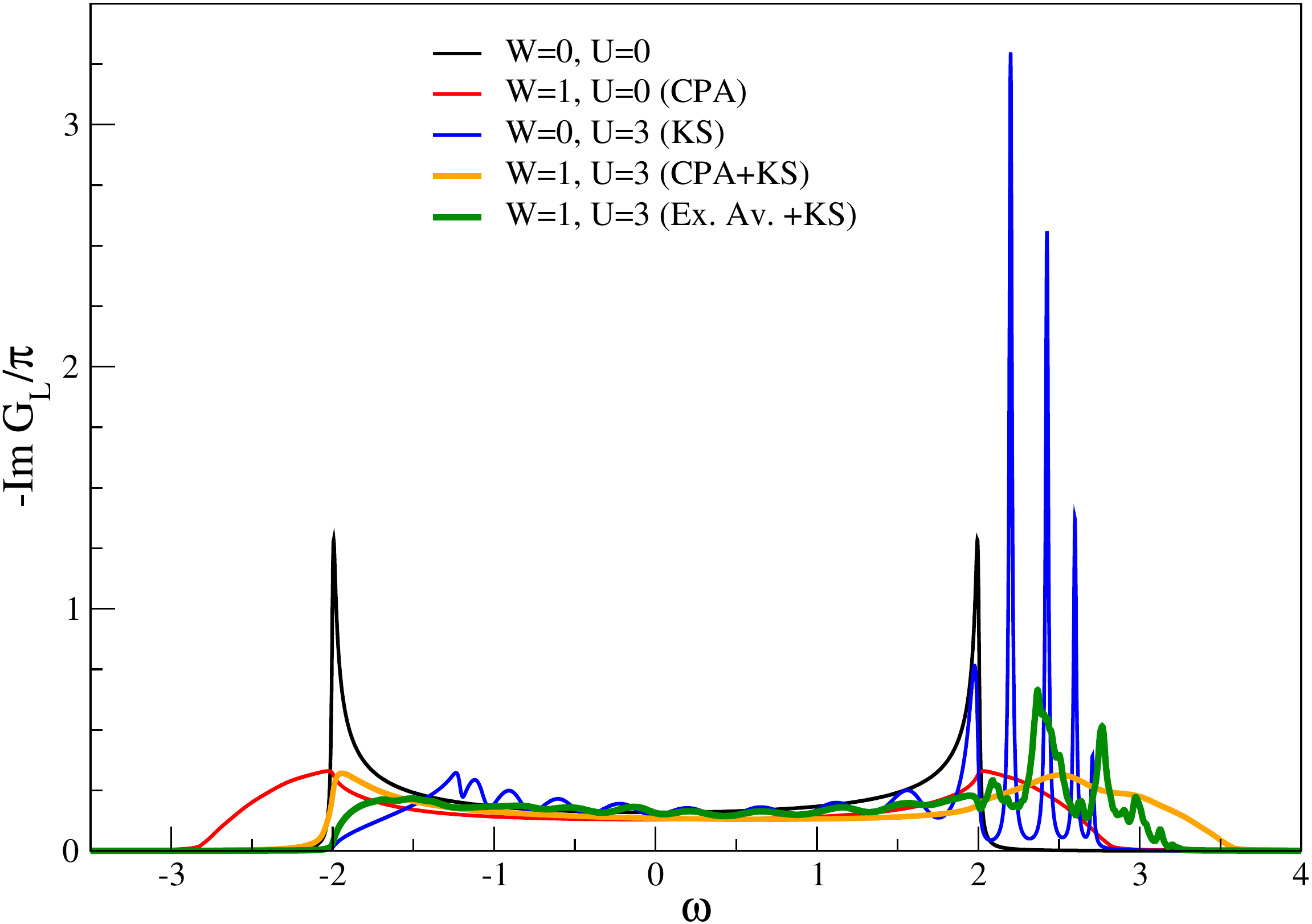}
\caption{\label{CPAfig2} (Color online) LDOS at an edge site for a disordered ($A_{50}B_{50}$) and 
interacting 14-site chain
contacted to two homogenous 1D leads at half-filling. The interaction $U=3$ and the disorder strength $W=1$.
The other system parameters are the same as in Fig.
\ref{CPAfig1}. The meaning of each curve is given by the color coding in the figure.
Also, {\it Ex. Av.} means complete numerical averaging, while {\it KS} indicates
that interactions are treated within a Kohn-Sham scheme.}
\end{figure}
For a short disordered chain connected to homogeneous leads, a treatment based on the single-site CPA 
amounts to introducing a complex, energy- and site-dependent self-energy. Numerically,
this CPA procedure is quite convenient, since it amounts to solving self-consistently a set of equations.
But how accurate is the CPA for the kind of (quantum-transport) geometries considered in this paper? To answer
this question, we have considered a disordered chain with $L=14$ sites, in absence of particle-particle
interactions ($U=0$). The chain represent a $A_{25}B_{75}$ and $A_{50}B_{50}$ system, i.e., in the notation 
of Section \ref{CPAsection}, $c_A=0.5$ and a $c_A=0.25$, with $c_B=1-c_A$. We considered two disorder strengths, i.e. $W=\epsilon_A=-\epsilon_B$ equal
to $0.5$ and $1.0$. The quantity that we intend to examine is the KS average local density of states (LDOS),
defined as $\langle d(\omega) \rangle_{ii}=-\pi^{-1}\Im \langle \hat{G}_{KS}(\omega)\rangle_{ii}$ ($i$
labels the i-th site in the chain), with $\langle \hat{G}_{KS}(\omega)$ obtained as described in 
Section \ref{CPAsection}. Strictly speaking, the LDOS is not accessible to ground state DFT; 
nevertheless, we think is instructive to look at this quantity in the framework of DFT,
to compare CPA versus exact disorder averaging when interactions are present.
For numerical convenience, the calculations were performed with an additional small Lorentzian broadening. 
The results for $U=0$ are presented in Fig. \ref{CPAfig1} (see the figure caption
for additional details). In each panel, the LDOS:s are obtained with both exact averages, summing
over $\binom{14 }{7}=3432$ configurations, and with the CPA. Furthermore, in each panel, we show LDOS:s at sites 
adjacent to the leads (labeled as "edge") and at a site in the center of the chain
(labeled as "middle"). Irrespective of the strength of the disorder, the results show that, overall,
the CPA (at least in this simple single-site formulation) provides a fair account of the role of the disorder, 
but much of the sharp structures in the exact curves are washed out. 
For example, for $W=1$, 
for $\omega \gtrsim 1$, we note a significant depression of the exact LDOS, which is completely
missed by the CPA. More in general, the sharp structures (bound states) outside the continuum are removed
by the CPA. This can have consequences in the long-time limit of quantum transport:
for pure systems, bound-states in the final state KS Hamiltonian can give rise to steady-state oscillations \cite{GLbound, Elham}, but, due to disorder as treated in the CPA, such long-lived oscillations are expected to be damped \cite{checkgamma}.
The situation is less clear for the exactly averaged LDOS: one can expect a self-averaging of the current and/or 
density oscillations, when the chain increases in size. However, for short chains, and for the
simple type of binary, on-site disorder considered here, (small) long-time oscillations could
persist.
It can be of interest to see at which extent this behavior is modified in the presence of inter-particle interactions.
In Figure \ref{CPAfig2}, the LDOS:s for $U=0$ and $U=3$ are plotted for a chain with and without disorder.
The curves for $U=3$ were obtained with lattice DFT in the LDA.
Starting with the $W=0$ case, we note that an important the effect of the interactions are resonant structures
at the top of the band ($\omega \gtrsim 2$, blue curve). The addition
of disorder within a CPA treatment has an overall effect similar to what observed in Fig. \ref{CPAfig1}, namely both
the KS sharp structures are dramatically smoothened in the KS+CPA LDOS:s (orange curve) \cite{HFresults}.
As for the non-interacting case [for reference, the CPA result for $U=0$ is also
shown (red curve)], the use of the CPA could
considerably affect the long-time behavior of densities and currents induced by a bias.
This is because the sharp structures due to the interaction in the DFT-LDA (that could induce long time limit oscillations in an ALDA treatment) are smoothed by the CPA. For the exactly averaged interacting LDOS, we equally observe a clear broadening/reduction of the split-off KS structures, albeit less pronounced than in the CPA-KS curve. Due to the  artificial broadening we introduced in our calculations, it cannot be excluded that for short
chains the density and current oscillations would stay long lived. The results shown here
were obtained from the initial state Hamiltonian. However, similar split-off structure are present in the case of the final state Hamiltonian, and the actual oscillations, independently from the presence of disorder, would likely be
absent if memory effects were taken into consideration \cite{KBEALDAtransport11}.

%
\subsection{(TD) DFT results for short chains attached to leads: static, transient and steady-state regimes}
In general, for disordered chains attached to semi-infinite leads, an exact numerical treatment analogous
to the one which we will discuss in Sect.\ref{res_finitesystems} is not available. In this case, two suitable methods are
(TD)DFT and the Green's functions technique. These two methods are both exact in principle but,
in practice, the many-body self-energy in a propagator approach and $v_{xc}$ in TDDFT
are not known exactly, and approximations are in order. Here we consider a TDDFT description,
as presented earlier in the paper. Our only (but important) approximation will be the use of a local
density approximation (LDA) in the ground state, and its adiabatic counterpart (ALDA) during the
dynamics. The quality of these approximations in the present contexts will be briefly discussed
at the end of this section.

\subsubsection{In equilibrium: The inverse participation ratio.}
\begin{table}[b]
\centering
\topcaption{Inverse participation ratio in the ground state} %
\begin{tabular}{ccccccccc}
\toprule %
\multicolumn {1}{c}{} &$ n $ &$$& $ U=0$& & $U=2$ && $U=4$\\ \hline\hline
\multirow{3}*{$^{W=0}$}
& 1.0&  & $0.0\overline{6}$ &&  0.067 && 0.067       \\ 
& 0.5  &  & $0.0\overline{6}$  && 0.067 && 0.067         \\ \hline%
\multirow{3}*{$^{W=1}$}
& 1.0 &  &0.068  &&0.067 && 0.067 \\ 
& 0.5  &  & 0.072 && 0.071 && 0.069 \\ \hline
\multirow{3}*{$^{W=3}$}
& 1.0 &  & 0.077 && 0.073 && 0.072  \\ 
& 0.5  &  & 0.113  &&  0.096  &&  0.09 \\ 
\end{tabular}
  \label{IPRtable}
\end{table}
We begin studying the system in equilibrium. The central region is a chain of $L=15$ sites.
We consider three strengths of disorder in the chain, $W=0,1,3$ and three values of the interaction,
$U=0,2,4$. Both $U$ and $W$ are given in units of the hopping parameter. In Table \ref{IPRtable},
we show the results for the configuration-averaged IPR for different densities $n=n_\uparrow+n_\downarrow$ 
in the leads (quarter and half-filling cases). The average was performed over  20 configurations.
As discussed in Sect. \ref{IPRproposal}, the IPR is calculated in terms of the KS densities in the central 
region [Eq.(\ref{IPRUS2})]. 

\begin{figure*}[tbhc]
\includegraphics[width=17.5cm]{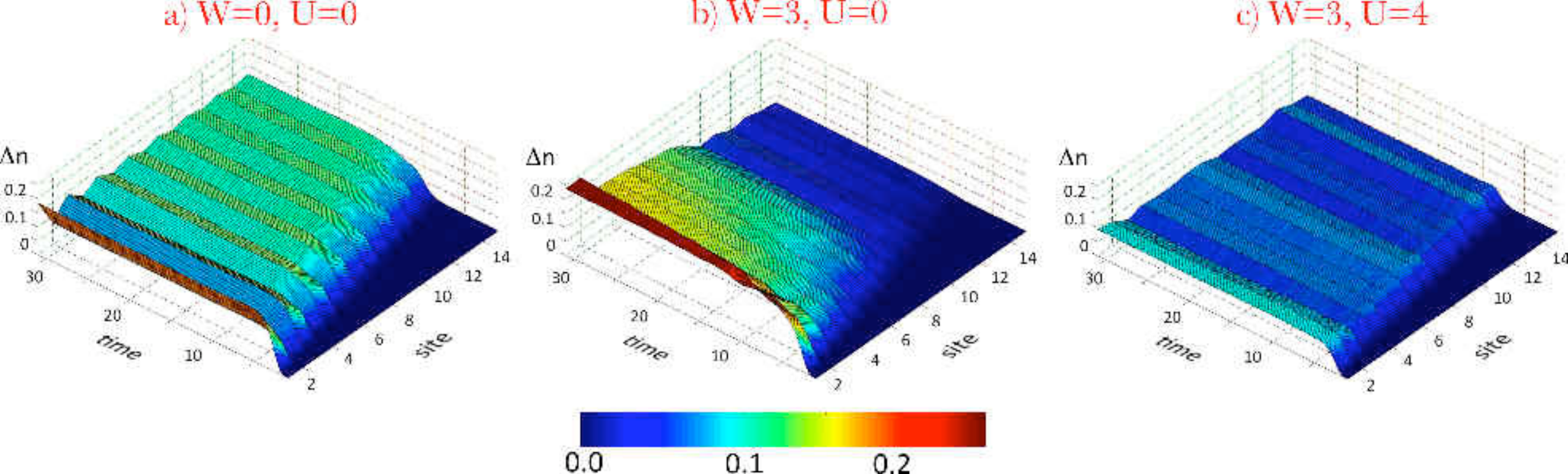}
\caption{\label{3Dplots} (Color online) Difference between non-equilibrium and 
ground-state (initial) densities, in a chain with $L=15, n=1$ and box disorder $W$, when a bias $b_S(t)$ is applied in one of the two leads, and $b_0=1.5$ [see Eq.(\ref{biashape})].
The maximum evolution time is ${T_{max}=30}$, in units of the inverse hopping parameter in the chain.
\label{Fig3Dplot}
}
\end{figure*}

For $U=W=0$, the system is homogenous, and we have only fully delocalized states and, since $L=15$, we get
an $IPR_C=L^{-1}=0.0\overline{6}$. Accordingly, values larger than $0.0\overline{6}$ would denote a tendency to localization.
This is what we observe on moving to larger $W$ values, while keeping $U=0$: localization is maximal
for $W=3$, both at quarter and half filling (however, the degree of localization is different for the two fillings).
A similar dependence of the IPR on $W$ is observed for $U=2$ and $U=4$. However, a different behavior
is noted when changing $U$ at a fixed $W$ (i.e., moving horizontally in the table). We see that the IPR
stays approximatively constant  at low $W$ but, for larger disorder, the IPR decreases on increasing $U$
(see especially the case of $W=3$). That is, increasing $U$ decreases localization, a manifestation
of the competing behavior of interactions and disorder. This has been noted before for finite samples 
(e.g. in terms of exact diagonalization \cite{itdepends} or DMRG calculations 
\cite{juliawern}).
When $W$ and $U$ become both very large, calculations as in 
\cite{itdepends,juliawern} suggest that localization prevails. From our results, this should happen 
a $U$ values larger than those in the table (e.g. for $W=3$, at $U > 4$). However, for such interaction strengths,
the shortcomings of the Bethe-Ansatz LDA can become particularly severe \cite{CapelleMagnetic}.
Nevertheless, it is quite interesting that a competing regime between disorder and interactions 
is accounted for within our lattice DFT-LDA approach, and with disorder occurring 
only in a sub-region of the system.

\subsubsection{Time-dependent densities.}
This behavior should also manifest in the dynamical properties
of the chain. To see this, we studied the time evolution of the system,
after the application of a bias in the leads. Our choice was to apply the bias only in the left lead
[i.e., at all times, $b_D(\tau)=0$, see Eq.(\ref{generalH})],  with the following time dependence:
\begin{eqnarray}
b_S(\tau)= \left\{ 
  \begin{array}{l l}
    b_0 [1+\cos\pi(1+\frac{\tau}{T})]/2 &  \text{, \quad $0\leq \tau\leq T$}\\
    b_0 &  \text{, \quad$\tau > T$},\\
  \end{array} \right.
  \label{biashape}
\end{eqnarray}
where $T=3$ (all time quantities are expressed in units of the reciprocal chain hopping).
This choice of $T$ is somewhat arbitrary, but in this way the effect of $b_S(\tau)$ is 
rather gradual, a situation expected to be favorable to the use of an ALDA based on the Bethe-Ansatz
for the 1D Hubbard model.
Our time-dependent results 
were obtained with propagation time steps of either $\Delta \tau= 
0.0025$ or $\Delta\tau= 0.0050$ and, as in the static case, averaged 
over 20 configurations. 

In Fig. \ref{Fig3Dplot}, we show the effect of disorder and interactions on the time-dependent density,
when $b_S(t)$ has been applied, with $b_0=1.5$. The chemical potential in the system was chosen 
to have half-filling in the leads (i.e. $n=n_\uparrow+n_\downarrow=1$). 
For convenience, we show the deviation of the density
$\Delta n_i=n_i(\tau)-n_i(0)$, rather than the density itself, since it illustrates more directly the
changes in the system. The left panel of Fig. \ref{Fig3Dplot} corresponds to when neither disorder nor
interactions are present in the chain.
In the transient phase, for sites close to the biased lead, we observe a quite sharp rise of the density, 
while the change in density occurs more smoothly for sites
closer to the unbiased lead. 
It is also clear that the densities in the chain attain a steady-state value
rather soon, already at $\tau \lesssim 15$, and that there is a quite regular propagation of the density front
across the chain. Disorder modifies in a quite substantially way the situation just described. In Fig. \ref{Fig3Dplot}b),
we note an increased $\Delta n$ for sites close to the biased lead, but the profile of the density propagation
front is now more irregular and significantly
attenuated inside the chain. This is also observed for weaker disorder ($W=1$, not shown), although
the differences from the homogeneous case are smaller. 

An interesting fact occurs when introducing
interactions ($U=4$) in the chain (Fig. \ref{Fig3Dplot}c). Now the time-dependent density landscape
recovers much of the regularity of the $U=W=0$ case, and the propagation of the density wave is
considerably less attenuated (with respect to Fig. \ref{Fig3Dplot}a,
the values of $\Delta n$ in the long-time limit are
reduced). So, it appears that even in the time-dependent case, interactions
can reduce the effect of disorder. We already noted such competition of effects 
when discussing the IPR in the ground state, but the results of Fig. \ref{Fig3Dplot}, and other
cases we have analyzed, not shown here, confirm the robustness of this behavior
with respect to i) bias strength (we also considered $b_0=0.5$), ii) particle density
(we also investigated the quarter filling regime) and, of course, iii) disorder and/or 
interaction strength. 
\subsubsection{Time-dependent inverse participation ratio.}
\begin{figure}[tbh]
\includegraphics[width=8.5cm]{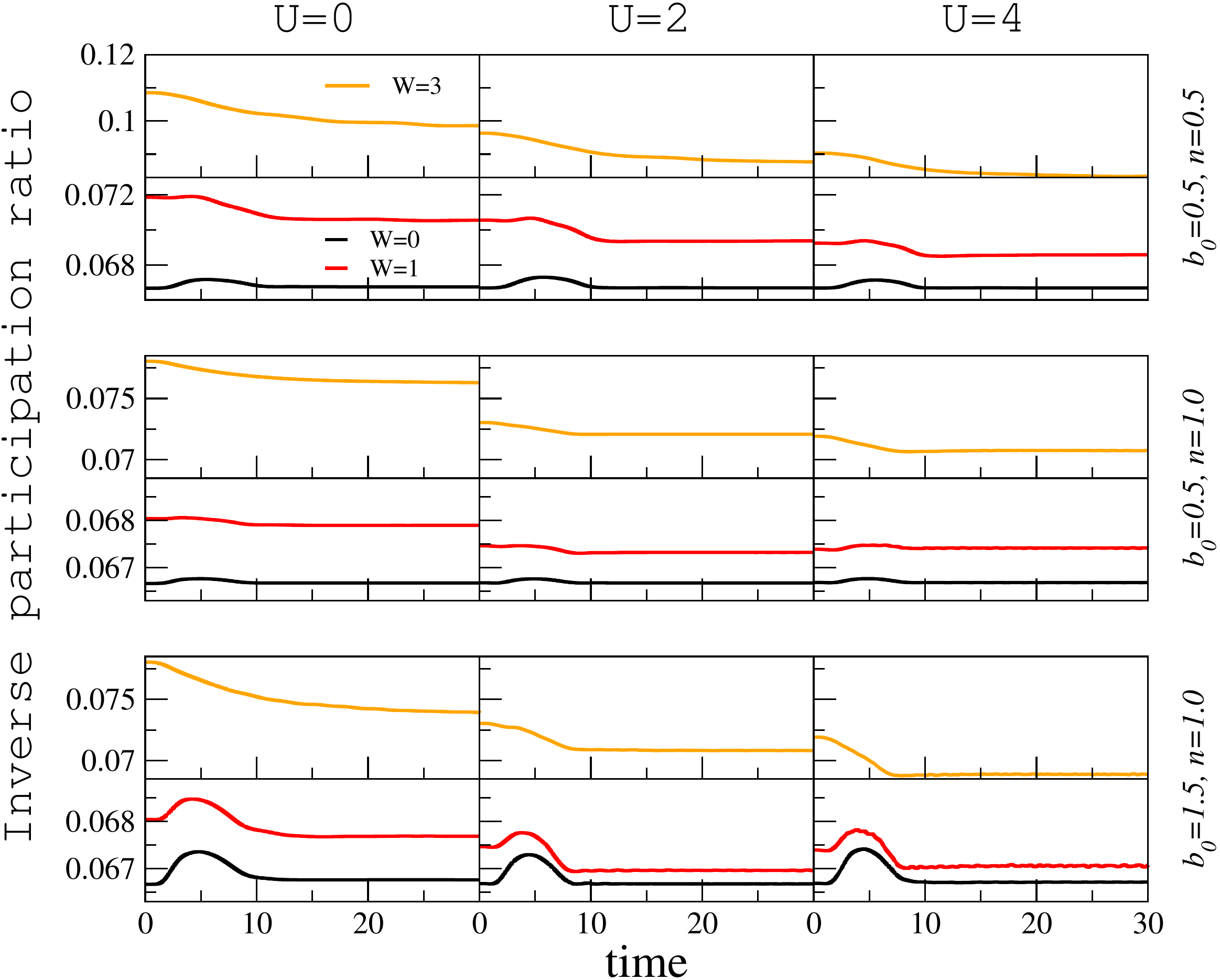}
\caption{ (Color online) Time-dependent IPR for a disordered and interacting chain with $L=15$ sites,
contacted to semi-infinite leads. Results are shown for different values of box disorder ($W$), interaction ($U$), bias
($b_0$) strengths, and for different lead densities $n$. Arithmetical disorder averaging is performed over 20 configurations.
Panels with one (orange) curve refer to the case of $W=3$, while those with two (black and red) curves refer to
$W=0$ and $W=1$, respectively. Panels in the same column pertain to same interaction value $U$, whilst 
panels in a row refer to common values of $b_0$ and $n$, as specified on the right of that row. All panels share the
same horizontal time interval, but scales on the vertical axes are different. \label{TDIPR}
}
\end{figure}
Also accessible within a Kohn-Sham lattice-TDDFT scheme is the time-dependent inverse participation ratio,
which we define via a simple modification of Eq. (\ref{IPRUS2}):
$IPR^{KS}_C(\tau) \equiv \sum_{i\in C} [n_i^{KS}]^2(\tau) / [\sum_{i\in C} n_i^{KS}(\tau)]^2$. Results for the time-dependent 
IPR are shown in Fig.\ref{TDIPR}
(see the figure caption for a definition of all the parameters). From Fig. \ref{TDIPR}, we observe that a larger disorder
induces a larger IPR, also in the dynamical regime. This holds for all cases examined in the figure; at the same time,
for a fixed disorder strength, interactions make the system more delocalized in time (as before, a complete
delocalization corresponds to an IPR $= 1/L=0.0\overline{6}$. At long times, the IPR is reduced
compared to its initial value;
such decrease is almost monotonic for large $W$, while at smaller disorder strengths
the IPR grows at first and then eventually becomes smaller. The region of increased IPR correspond to the transient phase,
where the variance among the different densities in the chain is largest (we have verified that the position at which the
IPR attains its maximum value depends on the way the bias is ramped-up). Conversely, the small IPR at long times
shows that, on average, the densities have the least mutual variance in a regime where a steady-state current can be attained.
\subsubsection{Time-dependent currents and the steady state regime.}
\begin{figure}[b]
\includegraphics[width=8.5cm]{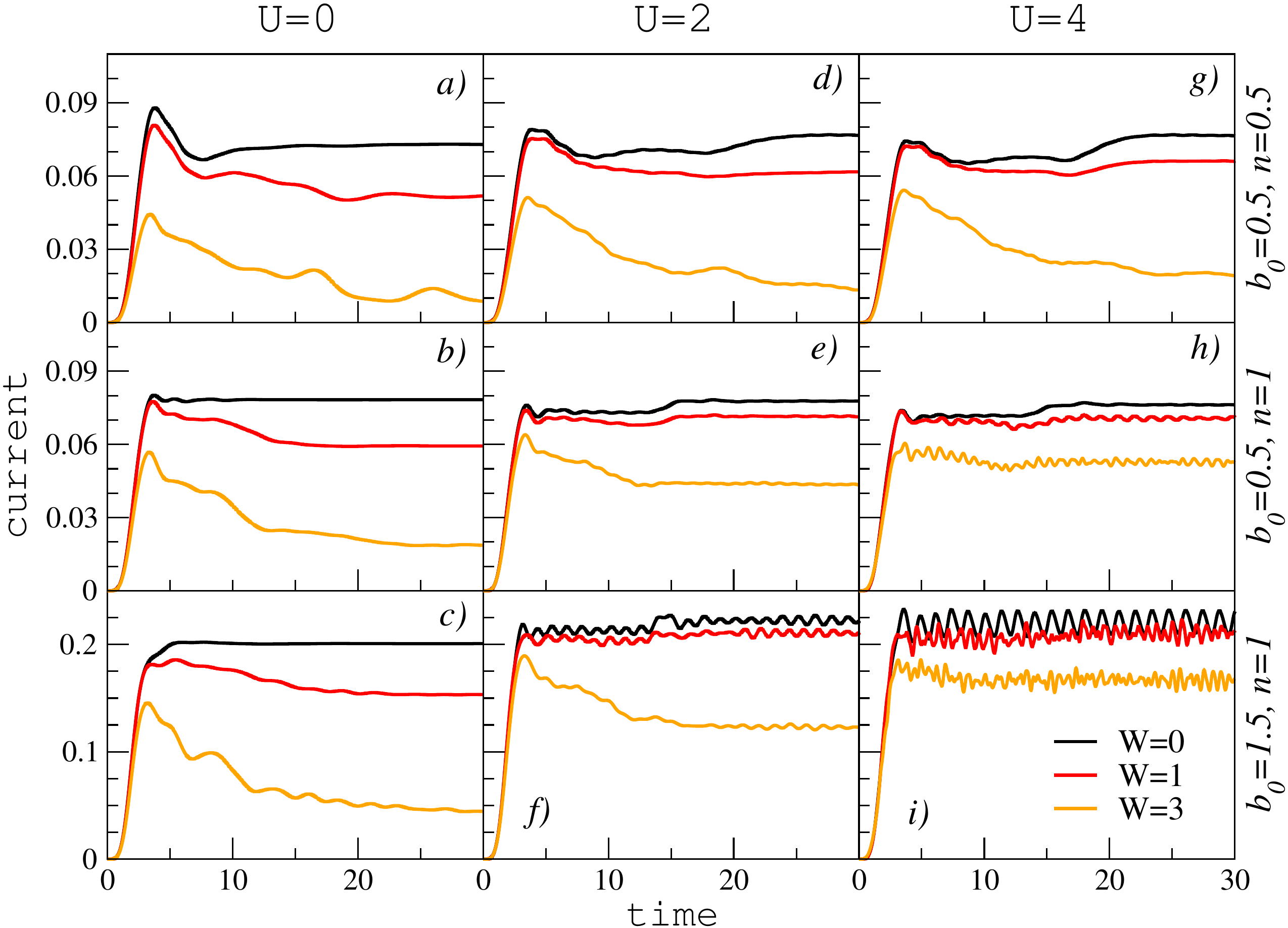}
\caption{\label{Thecurrents} (Color online) Time-dependent average currents for a chain with $L=15$ sites, 
and different interactions strengths ($U=0,2,4$). The currents shown are 
computed at the leftmost bond in the chain.
On the horizontal axis, the time is expressed in units of the 
inverse hopping parameter. The bias and the band filling 
for panels in the same row are specified on the right, while in each panel, the current 
is displayed for three disorder strengths  ($W=0,1,3$). Color coding for all panels is specified in the right lowermost panel. The configuration averages were obtained from 20 instances of box disorder.
}
\end{figure}
A more direct way to look at the competition of disorder and interactions in the dynamical regime
is analyzing the behavior of the current (in our case, the charge current). In Fig.\ref{Thecurrents}, we present results
for the average current at the leftmost-bond in the chain. Altogether, the different panels of Fig.\ref{Thecurrents}
show the current 
for several values of disorder/interaction strengths, of the biases, and the band filling (see the figure caption for
details on how the results are presented). We start with panel a),
corresponding to $U=0$, a bias $b_0=0.5$, and a density (in the leads) $n=0.5$ (quarter filling). 
For $W=0$ (black curve), the current reaches its steady-state value after a relatively short transient. 
On adding disorder ($W=1$, red curve), the length of transient increases, but eventually a steady state is reached.
On further increasing $W$, a steady-state current is not reached within the simulation
window (orange curve). However, the current in Fig.\ref{Thecurrents} (orange curve) and the currents at the other bonds in the chain (not-shown) become progressively close to the same average value (with smooth and decaying oscillations), suggesting that a steady state is reached outside our simulation window.
The overall trend in the panel a) is that an increasing disorder reduces the long-time (steady-state) value of the current.
Analogous behavior is seen in panels a), b) of the $U=0$ column, which correspond to different choices of
the bias $b_0$ and the density $n$ in the leads. Moving to the other two columns ($U=2,4$), we see that the changes
in each panel follow the same pattern, namely when disorder increases the current gets reduced.

A different perspective emerges from Fig. \ref{Thecurrents} when we mutually compare panels within the same row [e.g. panels a), d), c)].
In this case, for a given value of disorder strength, bias and density, the current increases at larger values of $U$,
an effect of competing disorder and interaction in the non-equilibrium regime. In the first row of Fig. \ref{Thecurrents} the current is always increasing when $U$ becomes larger, independently
of the value of $W$. Conversely, in the other two rows, depending on the value of $W$, the current can have a  non-monotonic 
dependence on $U$. Such non-monotonic (dynamical) behavior is consistent with results from ground-state
studies (see e.g. \cite{juliawern}), and its dependence on quantities such as the density $n$ or the disorder strength $W$
is plausible. To have a more complete picture of this tendency, calculations for larger $U$ values should be performed;
however, as mentioned earlier, for those $U$ values, the utility of an ALDA-TDDFT approach would
be significantly diminished.

So far, we have not mentioned at all the oscillatory behavior of the current in some of the panels of Fig. \ref{Thecurrents}.
Current and/or density oscillations in the long-time limit can result from different factors, such as single-particle bound
states \cite{GLbound,Elham} (in our case, due to diagonal disorder), discontinuities in the XC potential \cite{KSKVG10},
or a sloshing motion of the charge between different regions of the device and/or the terminal sites of the leads.
This latter mechanism has been pointed out and analyzed in a study \cite{PetriImagecharge} of the dynamical effects of image 
charge in quantum transport. In our system, we have observed that different and independent sets of oscillations can 
emerge in different parts of the chain, due to the inhomogeneities in the energetic landscape introduced by disorder
(current and cumulative density oscillations are more/less pronounced in different subregions of the chain).  
In recent work \cite{KBEALDAtransport11,VKPAB_ChemPhys11,PetriImagecharge}, Hartree-Fock and TDDFT-ALDA approaches
have been compared to results from Kadanoff-Baym dynamics. These studies clearly point out the importance of nonlocal
(in space and time) contribution beyond the instantaneous density, and suggest \cite{KBEALDAtransport11,PetriImagecharge} 
that current/density oscillations of the kind mentioned above are likely fragile against the inclusion of memory effects.
We wish to add here that, for disordered systems, another obvious cause of dephasing of the oscillations is disorder averaging.
This can be already argued at the non-interacting level. While bound-states can certainly be present in a specific instance of disorder, 
the induced oscillations are most probably to be washed out by configuration averaging, due to destructive interference among the oscillation
from different configurations. This receives indirect support from observing that already in the ground state of a 
device with binary disorder (Fig. \ref{CPAfig2}), split-off structures are largely reduced or washed out by exact or CPA averaging.
In our time-dependent calculations, as commonly done in the literature, configuration averages are based on a limited (and thus incomplete) number of random disorder realization; in this case the aforementioned cancellation effects of the oscillations can be incomplete. 
In any case, the feature emerging from our calculations, namely a competition between disorder and interaction,
appears to unrelated to current oscillations, since it present itself also when such oscillations are missing [see e.g., in panels a), d), g) in Fig. \ref{Thecurrents}].

As a final, but important remark, we observe that our results for the current lack of "reciprocity", i.e. for a fixed $U$, and the $W$:s considered,
the current is monotonically decreasing as a function of $W$, while a competing behavior could be expected from studies in the 
ground state \cite{juliawern}. 
This can possibly be due to a limitation of the ALDA (and not of TDDFT, which is in principle an exact theory). At the same time, we wish
to point out that, out of equilibrium and in the non-linear regime, the strength of the bias $b_0$  
can also significantly influence the competition between disorder and interactions (results for $b_0=1.5, n=0.5$, not shown here, 
compared to those in panels a),d),g), are consistent with this assertion). 

\subsection{Isolated short chains: TDDFT and exact results.}\label{res_finitesystems}
\begin{figure}[tbh]
\includegraphics[width=8.5cm]{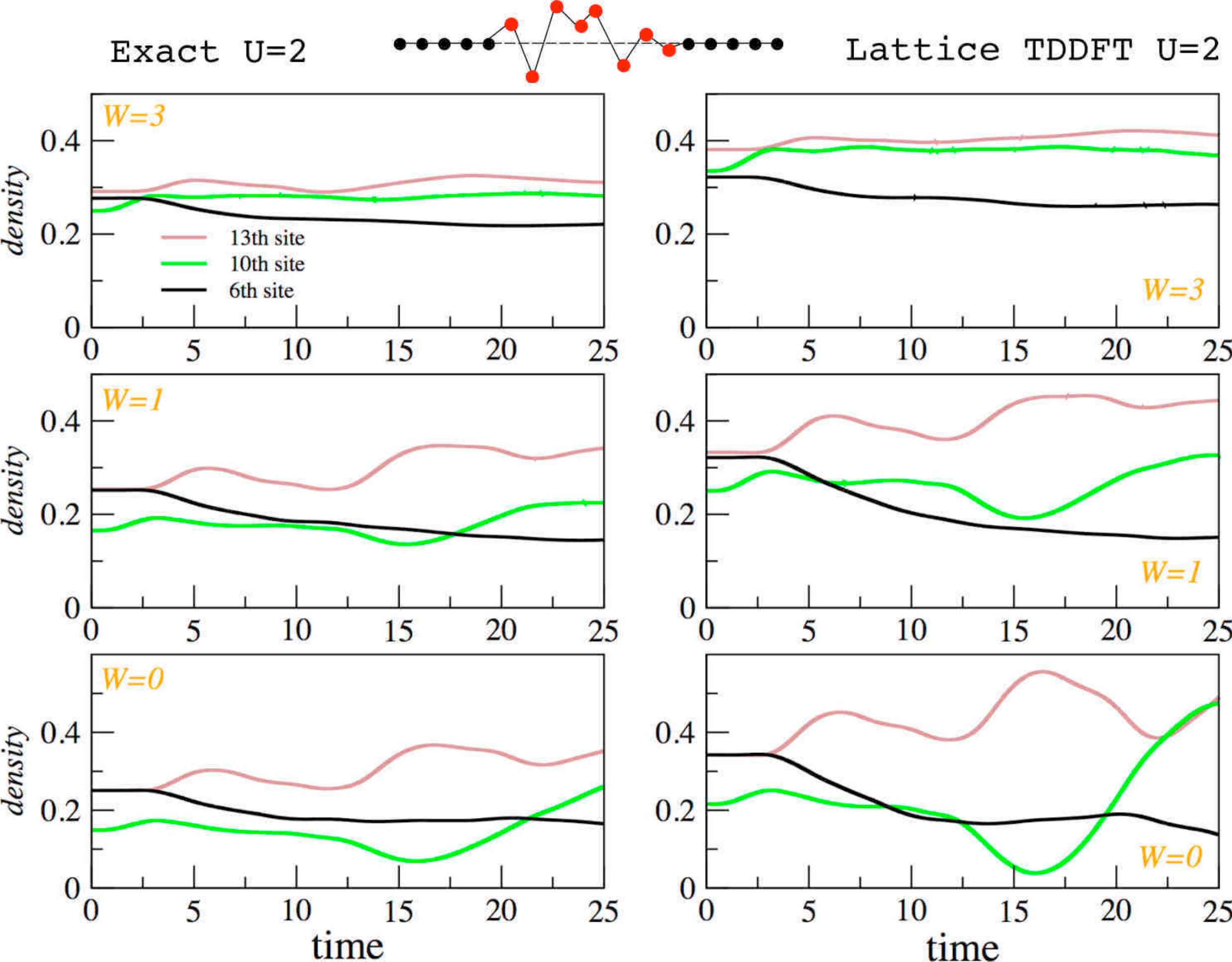}
\caption{\label{Cluster} (Color online) Exact (left panels) vs lattice TDDFT (right panels) averaged density results for an isolated chain with $L=18$
sites and $U=2$. The chain has a central region of 8 sites, with disorder and interaction, and two finite "leads" of
5 sites each (see the schematic rendering). In the chain there are $N=4$, spin-compensated particles and the dynamics
is induced with a time dependent "bias" which is uniformly applied to the nine-leftmost sites. The time
dependence of the bias is the same of Eq.(\ref{biashape}), with strength $b_0$=0.5. The color coding
in the top-left panel applies to all panels in the figure. The disorder averaged densities were obtained using
100 disorder configurations.
}
\end{figure}
The quantum transport results just discussed are based on approximate
XC potentials, obtained via an (adiabatic) local density approximation to $v_{xc}$. 
Comparisons between exact and (A)LDA results have been performed before in different contexts \cite{CapellePolini,Verdozzi08,DKAPCV11,VKPAB_ChemPhys11, KBEALDAtransport11},
but without considering the case of interest here, namely configuration averaged quantities
in disordered and interacting samples contacted to semi-infinite leads. 

An ideal way of doing this would be using time-dependent DMRG results as benchmark,
to obtain a comprehensive assessment of lattice (TD)DFT for the transient behavior of 
current and densities. From the computational point of view, this kind
of investigations are expected to be rather expensive, and we are not aware
of any published work on the subject. Here we take a much simpler view, and
consider small isolated disordered and isolated chains with few electrons, which can 
be treated via exact diagonalization. An extensive comparison between exact and lattice
(TD)DFT results is outside the scope of this work, and we briefly discuss just one example,
using a single case to gain some general insight.

In Fig. \ref{Cluster}, we examine the time-dependent densities, for a short isolated cluster
(all the parameters are specified in the figure caption). We mention that, differently from the previous sections, 
the "bias" is applied on the left half of the cluster, i.e. also in part of the interacting, disordered region 
(see the schematic rendering in at the top of Fig. \ref{Cluster}). Also, 
due to the small size of our system,
several reflections at the cluster boundary occur during the simulation interval considered
in Fig.\ref{Cluster}. Without going in any detail, the salient features emerging from
Fig.\ref{Cluster} are that TDDFT performs better at larger disorder strength, 
and that, compared to the exact results, TDDFT enhances the spread
among the time-evolved densities. From this example, it is apparent that lattice (TD)DFT is able to reproduce the
qualitative behavior of the exact results, but good quantitative agreement is lacking.
More investigations in this direction and an improvement of the XC potentials are certainly required. 
%
\section{Conclusions}\label{Conclude}
We have used lattice TDDFT to study the quantum transport properties of short, disordered and interacting
chains contacted to semi-infinite leads. Our work is largely exploratory in character, since we have
addressed only superficially several issues connected to the non-equilibrium physics of disordered 
interacting systems. 

In principle TDDFT is an exact approach, but (at times severe) approximations are 
usually made for the key quantity of the approach, the exchange-correlation potential. We have employed one
of them, the adiabatic local density approximation (ALDA).  Disordered systems, with a strongly varying local 
environment, are quite difficult tests for the ALDA which, however, by comparison to exact benchmarks,
certainly appears appropriate at the qualitative level for not too-fast varying time-dependent perturbations.

Within these boundaries, we have been able to address a type of system which is not easily accessible, for a reason
or another (limitations in principle, numerical costs, dimensionality, etc.) to several of the other methods
currently available. In fact, we are not aware of any existing work for the quantum transport geometries considered here,
where, in the presence of semi-infinite contacts, currents and densities in the disordered and correlated sample
have been followed in time from the initial transient phase to the long-time (possibly steady-state) regime.
To perform our study, we have introduced some modifications to the formalism, and modified the definition
of one of the standard indicators of localization, the inverse participation ratio. We have also explored the performance
of the coherent potential approximation, one of the popular schemes to perform disorder averaging.
The sharp spectral features due to disorder in the device are exceedingly smoothed by the CPA, and this can 
affect the behavior of the averaged steady-state currents.

Our time-dependent results show rather evident signatures of dynamical delocalization due to the
dynamical competition of disorder and interaction in the sample. This is consistent with the qualitative picture usually 
adopted for systems in the ground state, where interactions produce a "screening" of the disorder - i.e., the 
"attractive" behavior of low-energy impurities is compensated by the local repulsive interactions,
thus providing a less corrugated energy landscape. However, in the present case, an additional role is played
by the (non-weak) electric bias in the lead(s), which significantly (and dynamically) modifies the ground state energy landscape
and wavefunctions.  At which extent our findings remain robust against an improvement of the XC potentials is at present an open question.
This certainly calls for better potentials and future investigations \cite{KBEdisorder}, hopefully combined also with other methods.

We thank C.-O. Almbladh for valuable discussions and for critically reading the manuscript.
This work was supported by European Theoretical
Spectroscopy Facility  (INFRA-2007-211956).\newline
\section*{APPENDIX A: Lanczos time propagation \label{sec_appendixB}}
We briefly summarize the Lanczos method, as given in Ref. \onlinecite{JChemphys}. 
A useful comparative study between the Lanczos method and other integration schemes
can be found in Ref. \onlinecite{CastroMarques}.
Consider a system described by a TD Hamiltonian $H(t)$. If, for example,  we use the 
mid-point approximation for the time propagator and wish to evolve the system 
in the time interval $(t+\Delta, t)$, we obtain
\begin{equation}
|\Phi_{t+\Delta} \rangle=e^{-iH(t+\Delta/2)\Delta} | \Phi_t\rangle
\end{equation}
where $|\Phi_t\rangle$ is the (known) initial wavefunction.
Consider a finite Lanczos sequence $\{|V_k\rangle\}$, obtained by starting acting
on the  'seed'  $|\Phi_t\rangle\equiv |V_0\rangle $. Using $\{|V_k\rangle\}$ as a 
truncated basis, we get 
\begin{equation}
|\Phi_{t+\Delta}\rangle\approx\sum_{k=0}^{M_L} |V_k\rangle \; \langle V_k| e^{-iH_Lt} |V_0\rangle,\label{apross}
\end{equation}
where $H_L$ is the tridiagonal representation for $H(t+\Delta/2)$ in such a basis.
Inserting a complete set of eigenstates for the truncated space, 
 $H_L|\lambda\rangle=\epsilon_\lambda|\lambda\rangle$,
\begin{equation}
|\Phi_{t+\Delta}\rangle =\sum_{k=0}^{K} |V_k\rangle \; \left[\sum_\lambda  \langle V_k|\lambda\rangle  e^{-i \epsilon_{\lambda} t} \langle\lambda |V_0\rangle\right],
\end{equation}
where $|\Psi_{t+\Delta}\rangle$ is finally expressed in the basis of the original many body Hamiltonian.
The method requires a partial orthogonalization on the fly of the Lanczos basis in order to preserve accuracy along the trajectory.  For a simple estimate of the truncation error in Eq.(\ref{apross}), see the discussion in Ref. \onlinecite{JChemphys}.
\section*{APPENDIX B: Computationally efficient Lorentzian sums \label{sec_appendixA}}
To determine the energy dependent IPR of a finite system using the $N_\lambda$ eigenstates/values of the Hamiltonian,
we have to determine the local density of states (LDOS)  $n(\omega)=\sum_\lambda c^{\lambda}\delta(\omega-\epsilon_\lambda)$. This has to be
done for the $N_\omega$ values of the chosen energy grid. Furthermore, in some cases (as in our work here)
the IPR is also averaged over $N_D$ disorder configurations. 
Since the system is finite, it is expedient to introduce a Lorentzian broadening $\Gamma$, and to define a broadened LDOS
\begin{eqnarray}
n^\Gamma(\omega)\equiv n(\omega)\ast L^\Gamma(\omega)= \sum_\lambda c^\lambda L^\Gamma(\omega-\epsilon_\lambda),\label{broad}
\end{eqnarray} 
where $L^\Gamma(\omega)=(\Gamma/\pi)(\omega^2+\Gamma^2)^{-1}$.
When $N_\lambda, N_\omega, N_D$ are large, to compute $n^{\Gamma}(\omega)$ directly from Eq.(\ref{broad}) can be
computationally intensive. Fast Fourier Transform (FFT) is the method of choice in these cases,
but it requires uniform sampling, whilst the eigenvalues poles $\epsilon_{\lambda}$
are in general unevenly distributed.  This issue can be avoided with the approach described here, which is fast and accurate,
and in principle should be relevant to
FFT integration problems for large data set from nonuniform, adaptive or curvilinear sampling \cite{Jamie},.
To begin with, it is convenient to have a smooth function $C^{\gamma}(\omega)$ that decays rapidly in both 
$\omega$ and $t$ spaces, minimizes $\omega$ sampling and  allows smooth uniform sampling of
$ n^{C^{\gamma}}(\omega)=n(\omega)\ast C^{\gamma}(\omega)$. Then, the actual procedure is as follows:
i) we sample $n^{C^{\gamma}}(\omega),  L^{\Gamma}(\omega)$,  $C^{\gamma}(\omega)$; 
ii) via FFT, we compute $n^{C^{\gamma}}(t),L^{\Gamma}(t), C^{\gamma}(t)$; 
iii) we obtain $n^{\Gamma}(t)=[n^{C^{\gamma}}(t)/C^{\gamma}(t)] \times L^{\Gamma}(t)$;
iv) via inverse FFT, we compute $n^{\Gamma}(\omega)$.

The softening function $C^{\gamma}$ should be positive definite in both $\omega,t$ spaces, to ensure
easy deconvolution (we discard band limited functions such as rectangle, triangle, etc.,
since they require careful location of their zeroes). We also discard  
$C^{\gamma}=\exp(-\gamma\omega^{2})$, since a) $C^{\gamma}(t)$ may decay too fast, 
with deconvolution instabilities where $L^{\Gamma}(t)$ is still non negligible;
b) it is still expensive to determine $n(\omega) \ast \exp(-\gamma\omega^{2}) $ on $N_{\omega}$ 
sampling points. Based on these considerations, our optimal choice is 
\begin{equation}
\tilde{C}^{\gamma}(\omega)=\frac{\gamma}{4}(1-\gamma\frac{\partial }{\partial \gamma})e^{-\gamma
|\omega|}\propto e^{-\gamma|\omega|}\ast e^{-\gamma|\omega|}. 
\end{equation}
The function
$\tilde{C}^{\gamma}(\omega)$ decays exponentially, i.e. is "practically"
band limited with a small sampling domain; $\tilde{C}^{\gamma}(t)\propto (t^{2}+\gamma^{2})^{-2}$ is always positive,
and it decays slower than $L^{\Gamma}(t)$, thus avoiding deconvolution instabilities. Also, as results of
the self-convolution in the definition of $C^{\gamma}$,  the cusp $e^{-\gamma|\omega-\epsilon_{\lambda}|}$, when $\epsilon_{\lambda}$
is off-grid, is smoothed, reducing sampling errors. 
Finally, as a crucial advantage of the method, $n^{C^{\gamma}}(\omega)$ can be
computed recursively, needing the $\lambda$-sum just once (instead of at all the sampling points). In fact,
writing $n^{\tilde{C}^{\gamma}}(\omega) = \tilde{n}_{+}(\omega)+\tilde{n}_{-}(\omega)$, with 
$n_{\pm}(\omega) = \sum_{\lambda} c^{\lambda} e^{\pm\gamma(\omega-\epsilon_{\lambda})}
\Theta(\pm(\epsilon_{\lambda}-\omega))$, we get ($\Theta$ is the Heaviside function):
\begin{eqnarray}
\tilde{n}_{\pm}(\omega\mp \Delta)  = e^{-\gamma\Delta}[\tilde{n}_{\pm}(\omega)
+\gamma\Delta n_{\pm}(\omega)]\nonumber\\
+(1-\gamma\frac{\partial }{\partial \gamma})\times
\sum_{\lambda} c^{\lambda}
e^{\pm\gamma(\omega-\epsilon_{\lambda}\mp\Delta)}\nonumber\\
\times\; \Theta(\pm(\omega-\epsilon_{\lambda}))\Theta(\Delta\pm(\epsilon_{\lambda}-\omega
)).
\end{eqnarray}
%


\begin{thebibliography}{999}
\bibitem{equivalent} The two treatments  
are of course equivalent when no approximations are introduced.
\bibitem{selfexpress} In our case, $\sigma_L=\sigma_R$, with $2\sigma_R(x)=x-\text{sgn}(x)\sqrt{x^2-4}$
when $|x|>2$, and  $2\sigma_R(x)=x-i\sqrt{4-x^2}$ for $x \le 2$.
\bibitem{howgroundstate} For clarity, we mention that the initial Kohn-Sham state for our transport calculations
is not obtained in term of these equations, but with a different procedure, the one described in Ref. \onlinecite{practical}.
\bibitem{checkgamma} We have verified that the smooth character of the CPA LDOS persists for smaller
(but nonzero) values of the artificial broadening.
\bibitem{HFresults} With respect to an Hartree-Fock+CPA treatment, the KS+CPA curve would result shifted to lower energies, since, irrespective of disorder, the effective interaction is reduced by the XC term.
\bibitem{KBEdisorder} A study of disorder and correlation effects in quantum transport using the time-dependent Kadanoff-Baym equations is currently under way.
\end{thebibliography}
\end{document}